\documentclass[12pt]{article}
\usepackage{amsmath,color}

\usepackage{amssymb}
\usepackage{bm}
\usepackage{braket}
\usepackage[dvips]{graphicx}
\usepackage{comment}

\begin{document}

\baselineskip=17pt

\begin{titlepage}
\rightline{\tt KUNS-3044}
\rightline\today
\begin{center}
\vskip 1.5cm
\baselineskip=22pt
{\Large \bf {The LSZ reduction formula from homotopy algebras}}\\
\end{center}
\begin{center}
\vskip 1.0cm
{\large Keisuke Konosu,${}^1$ Yuji Okawa,${}^1$}\\
\vskip 0.1cm
{\large Shono Shibuya${}^{\, 2}$ and Jojiro Totsuka-Yoshinaka${}^{\, 3}$}
\vskip 1.0cm
{\it {${}^1$ Graduate School of Arts and Sciences, The University of Tokyo}}\\
{\it {3-8-1 Komaba, Meguro-ku, Tokyo 153-8902, Japan}}\\
konosu-keisuke@g.ecc.u-tokyo.ac.jp,
okawa@g.ecc.u-tokyo.ac.jp
\vskip 0.2cm

{\it {${}^2$ Department of Physics, Nagoya University, Nagoya, Aichi 464-8602, Japan}}\\
shibuya.shono.n8@s.mail.nagoya-u.ac.jp
\vskip 0.2cm

{\it {${}^3$ Department of Physics, Kyoto University, Kyoto 606-8502, Japan}}\\
george.yoshinaka@gauge.scphys.kyoto-u.ac.jp
\vskip 2.0cm

{\bf Abstract}
\end{center}

\noindent
When we describe string field theory or quantum field theory in terms of homotopy algebras,
on-shell scattering amplitudes at the tree level are obtained
by the formula based on the minimal model.
While this formula can be extended to loop amplitudes
and it generates the correct set of Feynman diagrams,
the evaluation of each Feynman diagram may fail to be well defined
because of mass renormalization.
Furthermore, this formula does not explain why we use Feynman propagators in loops.
In this paper we first present the LSZ reduction formula in terms of quantum $A_\infty$ algebras,
which provides a well-defined prescription for loop amplitudes.
We then present a formula for connected correlation functions
based on quantum $A_\infty$ algebras,
and we use it to discuss the relation between the LSZ reduction formula
and the extension of the minimal model to loop amplitudes. 

\end{titlepage}


\section{Introduction}
\label{section-introduction}
\setcounter{equation}{0}

In string theory,
tree-level scattering amplitudes are calculated
by integrating correlation functions
of on-shell vertex operators
over the moduli space of Riemann surfaces.
In string field theory,
tree-level scattering amplitudes are expressed
in terms of propagators, string products, and external on-shell states.
Let us consider the structure behind this description of tree-level scattering amplitudes.

When we construct gauge-invariant actions of string field theory,
homotopy algebras such as
$A_\infty$ algebras \cite{Stasheff:I, Stasheff:II, Getzler-Jones, Markl, Penkava:1994mu, Gaberdiel:1997ia}
for open string field theory
or $L_\infty$ algebras~\cite{Zwiebach:1992ie, Markl:1997bj} for closed string field theory
are useful.
When we describe the theory in terms of a homotopy algebra,
the information on the action is encoded
in a linear operator which we call ${\bf M}$
acting on a vector space which we denote by $T \mathcal{H}$
in the case of $A_\infty$ algebras.\footnote{
In the next section
we will explain ${\bf M}$ and $T \mathcal{H}$ in more detail,
but we consider that it is useful to present some equations
using ${\bf M}$ and other operators acting on $T \mathcal{H}$
in the introduction prior to the explanations in the next section.
}
We devide ${\bf M}$ as
\begin{equation}
{\bf M} = {\bf Q} +{\bm m} \,,
\end{equation}
where ${\bf Q}$ desribes the free part
and ${\bm m}$ describes the interactions.

When we classically integrate out part of the degrees of freedom,
homotopy algebras provide a universal description
of the resulting effective theory.
The degrees of freedom of the effective theory
can be characterized by a projection operator ${\bf P}$,
and the propagator for the degrees of freedom
that are integrated out is described by ${\bm h}$.
Then the action of the effective theory is described by
\begin{equation}
{\bf P} \, {\bf Q} \, {\bf P}
+{\bf P} \, {\bm m} \, \frac{1}{{\bf I} +{\bm h} \, {\bm m}} \, {\bf P} \,.
\label{classical-universal-formula}
\end{equation}
The generation of the effective theory from the original theory,
\begin{equation}
{\bf Q} +{\bm m}
\quad \to \quad
{\bf P} \, {\bf Q} \, {\bf P}
+{\bf P} \, {\bm m} \, \frac{1}{{\bf I} +{\bm h} \, {\bm m}} \, {\bf P} \,,
\end{equation}
is known as the homological perturbation lemma.

The scattering amplitudes correspond
to the projection onto on-shell states.
The theory we obtain by the projection onto the cohomology of the kinetic operator
is called the minimal model.
Scattering amplitudes at the tree level are obtained from the second term of~\eqref{classical-universal-formula}:
\begin{equation}
\pi_1 \, {\bf P} \, {\bm m} \, \frac{1}{{\bf I} +{\bm h} \, {\bm m}} \, {\bf P} \,,
\end{equation}
where $\pi_1$ is the projection operator onto the one-string sector of $T \mathcal{H}$.
This is one description of the structure behind tree-level scattering amplitudes in string field theory.
Actually, ordinary quantum field theory such as $\varphi^3$ theory
can also be described by homotopy algebras~\cite{Hohm:2017pnh, Jurco:2018sby, Nutzi:2018vkl, Arvanitakis:2019ald, Macrelli:2019afx, Jurco:2019yfd, Saemann:2020oyz, Bonezzi:2023xhn},
and then tree-level scattering amplitudes are given by the same formula.
The description in terms of homotopy algebras is quite universal.
The difference is the choice of the vecor space $T \mathcal{H}$
according to the degrees of freedom of the theory.
As we have seen, we have a satisfactory picture of tree-level scattering amplitudes
based on homotopy algebras. How about loop amplitudes?

When we integrate out part of the degrees of freedom quantum mechanically,
homotopy algebras again provide a universal description
of the resulting effective theory.
In homotopy algebras, quantization corresponds to the following modification:
\begin{equation}
{\bf M} \quad \to \quad {\bf M} +i \hbar \, {\bf U} \,,
\label{quantization}
\end{equation}
where the operator ${\bf U}$ acting on $T \mathcal{H}$ is determined
once we specify the degrees of freedom of the theory.
If we subtract the free part ${\bf Q}$ from~\eqref{quantization}, we have
\begin{equation}
{\bm m} \quad \to \quad {\bm m} +i \hbar \, {\bf U} \,,
\end{equation}
so we expect that loop amplitudes are described by
\begin{equation}
\pi_1 \, {\bf P} \, ( \, {\bm m} +i \hbar \, {\bf U} \, ) \,
\frac{1}{{\bf I} +{\bm h} \, ( \, {\bm m} +i \hbar \, {\bf U} \, )} \, {\bf P}
= \pi_1 \, {\bf P} \, {\bm m} \,
\frac{1}{{\bf I} +{\bm h} \, {\bm m} +i \hbar \, {\bm h} \, {\bf U}} \, {\bf P}
\label{loop-amplitude-formula}
\end{equation}
with ${\bf P}$ corresponding to the projection onto on-shell states at the tree level.
Here we simplified the expression using $\pi_1 \, {\bf P} \, {\bf U} = 0$,
which will be seen from the explanation in the next section.
In fact, the correct set of loop diagrams is generated from~\eqref{loop-amplitude-formula}.
When we set external states to be on shell, however,
loop diagrams may diverge
because some of the internal propagators can be forced to be on shell.
Furthermore, we use Feynman propagators in quantum field theory,
but momenta in loops are not uniquely determined by the momenta of the external states
and this does not seem to be related to the projection onto on-shell states.

String theory suffers from the same problem,
and the integration of correlation functions
of on-shell vertex operators over the moduli space of Riemann surfaces
no longer works for loop amplitudes when the mass is renormalized.
This problem can be solved by using string field theory.
It has been developed
how we should treat mass renormalization and vacuum shift
based on the one-particle irreducible (1PI) effective action.\footnote{
Mass renormalization and vacuum shift were originally discussed
in~\cite{Pius:2013sca, Pius:2014iaa, Pius:2014gza, Sen:2014dqa, Sen:2015hha, Sen:2015uoa},
and the discussion was later reorganized in the review~\cite{deLacroix:2017lif}.
See also subsection~9.3 of the recent review~\cite{Sen:2024nfd} for a concise explanation.}
Unfortunately, we have not yet figured out a satisfactory description
of the 1PI effective action in terms of homotopy algebras.
On the other hand, we have the formula for correlation functions
based on $A_\infty$ algebras~\cite{Okawa:2022sjf, Konosu:2023pal, Konosu:2023rkm},
which we explain in section~\ref{section-correlation-functions}.\footnote{
The description of global symmetries and their anomalies
based on this formula was developed recently in~\cite{Konosu:2024dpo}.
Nonperturbative correlation functions are discussed in~\cite{Konosu:2024zrq}.
}
We can thus deal with mass renormalization
using the LSZ reduction formula
as we do in quantum field theory.
We explain the description of the LSZ reduction formula
based on quantum $A_\infty$ algebras in section~\ref{section-LSZ}.
The resulting expression of loop amplitudes looks
rather different from the formula~\eqref{loop-amplitude-formula}.
One difference is that contributions from disconnected diagrams
are included in the LSZ reduction formula based on the correlation functions,
while only connected diagrams contribute in the formula~\eqref{loop-amplitude-formula}.
Motivated by this, we derive a formula for connected correlation functions
based on quantum $A_\infty$ algebras in section~\ref{section-connected}.
This formula is one of the main results of this paper.
Using this formula,
we then discuss the relation between the LSZ reduction formula
and~\eqref{loop-amplitude-formula} in section~\ref{section-relation}.
Section~\ref{section-conclusions} is devoted to conclusions and discussion.
The expression of the coproduct~\eqref{coproduct-formula} plays an important role
in obtaining the formula for connected correlation functions in section~\ref{section-connected}.
The derivation of~\eqref{coproduct-formula} is also one of the main results of this paper,
and we present its calculations in appendix~\ref{appendix-coproduct}.

\section{Correlation functions from $A_\infty$ algebras}
\label{section-correlation-functions}
\setcounter{equation}{0}

In this section we briefly explain the formula for correlation functions
based on $A_\infty$ algebras~\cite{Okawa:2022sjf, Konosu:2023pal, Konosu:2023rkm}.
We use the string-field-theory-like description developed in~\cite{Konosu:2023pal}.
See this paper for further details.

When we describe a theory without gauge symmetries
or a theory after gauge fixing in terms of an $A_\infty$ algebra,
the degrees of freedom are specified by a vector space which we denote by $\mathcal{H}_1$.
For a scalar field $\varphi (x)$ in $d$ dimensions,
$\Phi$ in $\mathcal{H}_1$ is expanded as
\begin{equation}
\Phi = \int d^d x \, \varphi (x) \, c(x) \,,
\end{equation}
where the basis vector $c(x)$ is degree even.
For a Dirac field $\Psi_\alpha (x)$ in $d$ dimensions,
$\Phi$ in $\mathcal{H}_1$ is expanded as
\begin{equation}
\Phi = \int d^d x \, ( \, \overline{\theta}_\alpha (x) \, \Psi_\alpha (x)
+\overline{\Psi}_\alpha (x) \, \theta_\alpha (x) \, ) \,,
\end{equation}
where the basis vector $\theta_\alpha (x)$ is degree odd.

To describe observables,
we also use the dual vector space of $\mathcal{H}_1$.
For this purpose, we use a vector space $\mathcal{H}_2$
and a symplectic form $\omega$.
We define $\mathcal{H}$ by
\begin{equation}
\mathcal{H} = \mathcal{H}_1 \oplus \mathcal{H}_2 \,.
\end{equation}
For the scalar field theory, $\widetilde{\Phi}$ in $\mathcal{H}_2$ is expanded as
\begin{equation}
\widetilde{\Phi} = \int d^d x \, f(x) \, d(x) \,,
\end{equation}
where the basis vector $d(x)$ is degree odd
and $f(x)$ specifies the observable.
The symplectic form $\omega$ is defined by
\begin{equation}
\biggl(
\begin{array}{cc}
\omega \, ( \, c(x_1) \,, c(x_2) \, ) & \omega \, ( \, c(x_1) \,, d(x_2) \, ) \\
\omega \, ( \, d(x_1) \,, c(x_2) \, ) & \omega \, ( \, d(x_1) \,, d(x_2) \, )
\end{array}
\biggr)
= \biggl(
\begin{array}{cc}
0 & \delta^d ( x_1-x_2 ) \\
{}-\delta^d ( x_1-x_2 ) & 0
\end{array}
\biggr) \,.
\end{equation}
Since
\begin{equation}
\omega \, ( \, \Phi \,, \widetilde{\Phi} \, )
= \int d^d x \, f(x) \, \varphi (x) \,,
\end{equation}
this describes the element of the dual space of $\mathcal{H}_1$
specified by $f(x)$ as follows:
\begin{equation}
\varphi (x) \to \int d^d x \, f(x) \, \varphi (x) \,.
\end{equation}

The action is written as
\begin{equation}
S = {}-\frac{1}{2} \, \omega \, ( \, \Phi \,,\, Q \, \Phi \, )
-\sum_{n=0}^\infty \, \frac{1}{n+1} \,
\omega \, ( \, \Phi \,,\, m_n \, ( \, \Phi \otimes \ldots \otimes \Phi  \, ) \, ) \,.
\end{equation}
The kinetic term is described by $Q$, which is a linear map from $\mathcal{H}_1$ to $\mathcal{H}_2$.
For example, the kinetic term of the scalar field theory is described by
\begin{equation}
Q \, c(x) = ( {}-\partial^2 +m^2 \, ) \, d (x) \,,
\end{equation}
where $m$ is the mass of the scalar field.
Interactions of $O( \Phi^{n+1} )$ are described by $m_n$,
which is a linear map from $\mathcal{H}_1^{\otimes n}$ to $\mathcal{H}_2$.
For example, the cubic interaction of $\varphi^3$ theory is described by
\begin{equation}
m_2 \, ( \, c(x_1) \otimes c(x_2) \, )
= {}-\frac{g}{2} \, \int d^d x \, \delta^d (x-x_1) \, \delta^d (x-x_2) \, d(x) \,,
\end{equation}
where $g$ is the coupling constant.

It is convenient to consider linear operators
acting on the vector space $T \mathcal{H}$ defined by
\begin{equation}
T \mathcal{H}
= \mathcal{H}^{\otimes 0} \, \oplus \mathcal{H}
\oplus \mathcal{H}^{\otimes 2} \oplus \mathcal{H}^{\otimes 3} \oplus  \ldots \,,
\end{equation}
where we also introduced the vector space $\mathcal{H}^{\otimes 0}$.
It is a one-dimensional vector space
given by multiplying a single basis vector {\bf 1} by complex numbers.
The vector {\bf 1} satisfies
\begin{equation}
{\bf 1} \otimes \Phi = \Phi \,, \qquad \Phi \otimes {\bf 1} = \Phi \,, \qquad
{\bf 1} \otimes {\bf 1} = {\bf 1}
\end{equation}
for any $\Phi$ in $\mathcal{H}$.
We denote the projection operator onto $\mathcal{H}^{\otimes n}$ by $\pi_n$.
For a map $D_n$ from $\mathcal{H}^{\otimes n}$ to $\mathcal{H}$
with $n = 0, 1, 2, \ldots$,
we define an associated operator ${\bm D}_n$ acting on $T \mathcal{H}$ as follows:
\begin{equation}
\begin{split}
{\bm D}_n \, \pi_m & = 0 \quad \text{for} \quad m < n \,, \\
{\bm D}_n \, \pi_n & = D_n \, \pi_n \,, \\
{\bm D}_n \, \pi_{n+1}
& = ( \, D_n \otimes {\mathbb I} +{\mathbb I} \otimes D_n \, ) \, \pi_{n+1} \,, \\
{\bm D}_n \, \pi_m
& = ( \, D_n \otimes {\mathbb I}^{\otimes (m-n)}
+\sum_{k=1}^{m-n-1} {\mathbb I}^{\otimes k} \otimes D_n \otimes {\mathbb I}^{\otimes (m-n-k)}
+{\mathbb I}^{\otimes (m-n)} \otimes D_n \, ) \, \pi_m \\
& \qquad \text{for} \quad m > n+1 \,.
\end{split}
\label{coderivation-form}
\end{equation}
Here and in what follows we denote the identity operator on $\mathcal{H}$ by ${\mathbb I}$.
An operator acting on $T \mathcal{H}$ of this form is called a {\it coderivation}.
In this paper, we use a coderivation ${\bf \Phi}$
associated with $\Phi$ in $\mathcal{H}$.
It is defined by\footnote{
This corresponds to ${\bm D}_0$ in~\eqref{coderivation-form}.
The action of ${\bm D}_0$ may be slightly difficult to understand.
See subsection~2.2 of~\cite{Okawa:2022sjf} for a further explanation.
}
\begin{equation}
\begin{split}
{\bf \Phi} \, {\bf 1} & = \Phi \,, \\
{\bf \Phi} \, \pi_1
& = ( \, \Phi \otimes {\mathbb I} +{\mathbb I} \otimes \Phi \, ) \, \pi_1 \,, \\
{\bf \Phi} \, \pi_m
& = ( \, \Phi \otimes {\mathbb I}^{\otimes m}
+\sum_{k=1}^{m-1} {\mathbb I}^{\otimes k} \otimes \Phi \otimes {\mathbb I}^{\otimes (m-k)}
+{\mathbb I}^{\otimes m} \otimes \Phi \, ) \, \pi_m \quad \text{for} \quad m > 1 \,.
\end{split}
\end{equation}

Coderivations can be characterized using a linear operation called {\it coproduct}
which maps $T \mathcal{H}$
to a tensor product of two copies of $T \mathcal{H}$
denoted as $T \mathcal{H} \otimes' T \mathcal{H}$.
See appendix A of~\cite{Erler:2015uba} for further details.
We use the symbol~$\otimes'$ to distinguish
the tensor product of $T \mathcal{H}$ from
the tensor product~$\otimes$ within $T \mathcal{H}$.
We denote the coproduct by $\triangle$,
and its actions on ${\bf 1}$, $A_1$, $A_1 \otimes A_2$,
and $A_1 \otimes A_2 \otimes A_3$ are given by
\begin{align}
\triangle \, {\bf 1} & = {\bf 1} \otimes' {\bf 1} \,, \\
\triangle \, A_1 & = {\bf 1} \otimes' A_1 +A_1 \otimes' {\bf 1} \,, \\ 
\triangle \, ( \, A_1 \otimes A_2 \, )
& = {\bf 1} \otimes' ( \, A_1 \otimes A_2 \, )
+A_1 \otimes' A_2
+( \, A_1 \otimes A_2 \, ) \otimes' {\bf 1} \,, \\ 
\triangle \, ( \, A_1 \otimes A_2 \otimes A_3 \, )
& = {\bf 1} \otimes' ( \, A_1 \otimes A_2 \otimes A_3 \, )
+A_1 \otimes' ( \, A_2 \otimes A_3 \, ) \nonumber \\
& \quad~
+( \, A_1 \otimes A_2 \, ) \otimes' A_3
+( \, A_1 \otimes A_2 \otimes A_3 \, ) \otimes' {\bf 1} \,.
\end{align}
The action of the coproduct
on $A_1 \otimes A_2 \otimes \ldots \otimes A_n$ for $n > 1$
is as follows:
\begin{equation}
\begin{split}
\triangle \, ( \, A_1 \otimes A_2 \otimes \ldots \otimes A_n \, )
& = {\bf 1} \otimes' ( \, A_1 \otimes A_2 \otimes \ldots \otimes A_n \, ) \\
& \quad~
+\sum_{k=1}^{n-1} ( \, A_1 \otimes A_2 \otimes \ldots \otimes A_k \, )
\otimes' ( \, A_{k+1} \otimes A_{k+2} \otimes \ldots \otimes A_n \, ) \\
& \quad~
+( \, A_1 \otimes A_2 \otimes \ldots \otimes A_n \, ) \otimes' {\bf 1} \,.
\end{split}
\end{equation}
Using the coproduct, a coderivation is defined in the following way.
A linear operator ${\bm D}$ on $T \mathcal{H}$
is a coderivation when it satisfies
\begin{equation}
\triangle \, {\bm D}
= ( \, {\bm D} \otimes' {\bf I} +{\bf I} \otimes' {\bm D} \, ) \, \triangle \,,
\label{coderivation-definition}
\end{equation}
where ${\bf I}$ is the identity operator on $T \mathcal{H}$.
We can show using~\eqref{coderivation-definition} that ${\bm D}$ is determined from $\pi_1 \, {\bm D}$.
See appendix~A of~\cite{Erler:2015uba} for a proof.

It is also convenient to introduce a linear operation
which maps $T \mathcal{H} \otimes'T \mathcal{H}$ to $T \mathcal{H}$.
We denote this operation by $\bigtriangledown$,
and it acts by simply replacing $\otimes'$ with $\otimes$.
For example, its action on
$A_1 \otimes A_2 \otimes' A_3 \otimes A_4 \otimes A_5$ is given by
\begin{equation}
\bigtriangledown \,
( \, A_1 \otimes A_2 \otimes' A_3 \otimes A_4 \otimes A_5 \, )
= A_1 \otimes A_2 \otimes A_3 \otimes A_4 \otimes A_5 \,.
\end{equation}
The main identity we will use in what follows is
\begin{equation}
\pi_{m+n} = \bigtriangledown \, ( \, \pi_m \otimes' \pi_n \, ) \, \triangle
\end{equation}
for any pair of $m$ and $n$.
It is straightforward to prove this identity from the definitions
of $\triangle$ and $\bigtriangledown$.

We define the coderivation ${\bf Q}$
associated with $Q$
and the coderivation ${\bm m}_n$
associated with $m_n$ for each $n$.
We then define ${\bm m}$ by
\begin{equation}
{\bm m} = \sum_{n=0}^\infty {\bm m}_n \,,
\end{equation}
and we define ${\bf M}$ by
\begin{equation}
{\bf M} = {\bf Q} +{\bm m} \,.
\end{equation}
When we consider gauge theories, the action described by the coderivation ${\bf M}$
is gauge invariant if ${\bf M}$ satisfies
\begin{equation}
{\bf M}^2 = 0 \,.
\end{equation}

When we consider projections onto subspaces of $\mathcal{H}$,
homotopy algebras have turned out to provide useful tools.
The projection onto on-shell states
describes on-shell scattering amplitudes~\cite{Kajiura:2003ax},
the projection onto the physical sector leads
to mapping between covariant and light-cone string field theories~\cite{Erler:2020beb},
and the projection onto the massless sector is relevant
for the low-energy effective
action~\cite{Sen:2016qap, Erbin:2020eyc, Koyama:2020qfb, Arvanitakis:2020rrk, Arvanitakis:2021ecw}.
We consider projections which commute with $Q$.
We denote the projection operator by $P$:
\begin{equation}
P^2 = P \,, \qquad P \, Q = Q \, P \,.
\end{equation}
We then promote $P$ on $\mathcal{H}$ to ${\bf P}$ on $T \mathcal{H}$ as follows:
\begin{equation}
\begin{split}
{\bf P} \, \pi_0  & = \pi_0 \,, \\
{\bf P} \, \pi_n & = P^{\otimes n} \, \pi_n
\end{split}
\end{equation}
for $n > 0$, where
\begin{equation}
P^{\otimes n} = \underbrace{ \, P \otimes P \otimes \ldots \otimes P \, }_n \,.
\end{equation}
The operator ${\bf P}$ is a {\it cohomomorphism}.
A cohomomorphism ${\bm F}$ is characterized in terms of the coproduct $\triangle$ as
\begin{equation}
\triangle \, {\bm F} = ( \, {\bm F} \otimes' {\bm F} \, ) \, \triangle \,,
\end{equation}
and it is uniquely specified by this relation and $\pi_1 \, {\bm F}$.
See appendix~A of~\cite{Erler:2015uba} for details.
The operators ${\bf Q}$ and ${\bf P}$ satisfy
\begin{equation}
{\bf P}^2 = {\bf P} \,, \qquad
{\bf Q} \, {\bf P} = {\bf P} \, {\bf Q} \,.
\end{equation}

A key ingredient is an operator $h$ satisfying
\begin{equation}
Q \, h +h \, Q = {\mathbb I}-P \,, \qquad
h \, P = 0 \,, \qquad
P \, h = 0 \,, \qquad
h^2 = 0 \,.
\end{equation}
It is called a {\it contracting homotopy}, and physically it describes
propagators associated with degrees of freedom which are integrated out.
We then promote $h$ on $\mathcal{H}$ to ${\bm h}$ on $T \mathcal{H}$ as follows:
\begin{equation}
\begin{split}
{\bm h} \, \pi_0 & = 0 \,, \\
{\bm h} \, \pi_1 & = h \, \pi_1 \,, \\
{\bm h} \, \pi_2
& = ( \, h \otimes P +{\mathbb I} \otimes h \, ) \, \pi_2 \,, \\
{\bm h} \, \pi_m
& = ( \, h \otimes P^{\otimes (m-1)}
+\sum_{k=1}^{m-2} {\mathbb I}^{\otimes k} \otimes h \otimes P^{\otimes (m-1-k)}
+{\mathbb I}^{\otimes (m-1)} \otimes h \, ) \, \pi_m
\end{split}
\end{equation}
for $m > 2$.
The operator ${\bm h}$ is not a coderivation
and instead satisfies the following relation:
\begin{equation}
\triangle \, {\bm h} = ( \, {\bm h} \otimes' {\bf P} +{\bf I} \otimes' {\bm h} \, ) \, \triangle \,.
\end{equation}
The relations involving $Q$, $P$, and $h$ are promoted
to the following relations
\begin{equation}
{\bf Q} \, {\bm h} +{\bm h} \, {\bf Q} = {\bf I}-{\bf P} \,, \qquad
{\bm h} \, {\bf P} = 0 \,, \qquad
{\bf P} \, {\bm h} = 0 \,, \qquad
{\bm h}^2 = 0 \,.
\end{equation}
The important point is that the theory after the projection inherits
the $A_\infty$ structure from the theory before the projection.
As we mentioned in the introduction, it is given by the homological perturbation lemma:
\begin{equation}
{\bf Q} +{\bm m} \quad \to \quad
{\bf P} \, {\bf Q} \, {\bf P}
+{\bf P} \, {\bm m} \, \frac{1}{{\bf I} +{\bm h} \, {\bm m}} \, {\bf P} \,.
\end{equation}

On-shell scattering amplitudes are described by the projection onto on-shell states.
In the case of the projection onto on-shell states,
${\bf P} \, {\bf Q} \, {\bf P}$ vanishes
and on-shell scattering amplitudes at the tree level can be calculated from
\begin{equation}
\pi_1 \, {\bf P} \, {\bm m} \, \frac{1}{{\bf I} +{\bm h} \, {\bm m}} \, {\bf P} \,.
\end{equation}
Our main focus in this paper is the generalization of this formula to loop amplitudes.
Loop diagrams for on-shell scattering amplitudes can be generated from
\begin{equation}
\pi_1 \, {\bf P} \, {\bm m} \, \frac{1}{{\bf I} +{\bm h} \, {\bm m} +i \hbar \, {\bm h} \, {\bf U}} \, {\bf P} \,.
\end{equation}
In the case of scalar field theories,
the operator ${\bf U}$ is given by 
\begin{equation}
{\bf U} = \int d^d x \, {\bm c} (x) \, {\bm d} (x) \,,
\end{equation}
where ${\bm c} (x)$  and ${\bm d} (x)$ are coderivations associated with $c(x)$ and $d(x)$, respectively.
The operator ${\bf U}$ consists of linear maps from $\mathcal{H}^{\otimes n}$
to $\mathcal{H}^{\otimes (n+2)}$ so that $\pi_1 \, {\bf U} = 0$.
It then follows that $\pi_1 \, {\bf P} \, {\bf U} = P \, \pi_1 \, {\bf U} = 0$,
and this was used to simplify~\eqref{loop-amplitude-formula} in the introduction.

When we consider correlation functions,
we carry out the path integral {\it completely}.
This should correspond to the projection with
\begin{equation}
P=0 \,.
\end{equation}
However, there are plane-wave solutions in Lorentzian theories,
and the cohomology of $Q$ is nontrivial
so that we cannot choose the projection with $P=0$.
The resolution is related
to how we define the vacuum state in the path integral.
In the case of scalar field theories,
we choose the asymptotic behavior of the scalar field $\varphi$
such that the imaginary part of $\varphi$ is infinitesimally positive
in the limit ${\rm Re} \, \varphi \to -\infty$
and infinitesimally negative in the limit ${\rm Re} \, \varphi \to \infty$.
This corresponds to the modification
\begin{equation}
Q \quad \to \quad Q_\epsilon
\quad
\text{with}
\quad
Q_\epsilon \,  c(x) = ( {}-\partial^2 +m^2 -i \epsilon \, ) \, d (x) \,,
\end{equation}
where the constant $\epsilon$ is infinitesimally positive.
The cohomology of $Q_\epsilon$ is then trivial,
and we can consider the projection with $P=0$.

The conditions for $h$ when $P=0$ are
\begin{equation}
Q_\epsilon \, h +h \, Q_\epsilon = {\mathbb I} \,, \qquad
h^2 = 0 \,.
\end{equation}
The operator $h$ satisfying these relations
can be constructed using the Feynman propagator.
This reproduces the perturbation theory in the path integral
where the Feynman propagator is used in loop diagrams.

The operator ${\bf P}$ when $P$ vanishes
corresponds to the projection onto $\mathcal{H}^{\otimes 0}$:
\begin{equation}
{\bf P} = \pi_0 \,.
\end{equation}
The operator $\bm{h}$ when $P$ vanishes is given by
\begin{equation}
\bm{h} = h \, \pi_1
+\sum_{n=2}^\infty ( \, \mathbb{I}^{\otimes (n-1)} \otimes h \, ) \, \pi_n \,.
\end{equation}
We claim that correlation functions are given by
\begin{equation}
\langle \, \Phi^{\otimes n} \, \rangle = \pi_n \, {\bm f} \, {\bf 1}
\end{equation}
with
\begin{equation}
\Phi^{\otimes n} = \underbrace{\, \Phi \otimes \Phi \otimes \ldots \otimes \Phi \,}_n \,,
\end{equation}
where ${\bm f}$ is
\begin{equation}
{\bm f} = \frac{1}{{\bf I} +{\bm h} \, {\bm m} +i \hbar \, {\bm h} \, {\bf U}} \,.
\label{f-definition}
\end{equation}
The correlation functions can be extracted by expanding $\langle \, \Phi^{\otimes n} \, \rangle$
in the basis of $\mathcal{H}_1^{\otimes n}$.
For example, the correlation functions of scalar field theories
$\langle \, \varphi (x_1) \, \varphi (x_2) \, \ldots \, \varphi (x_n) \, \rangle$
appear as coefficients in front of
$c (x_1) \otimes c(x_2) \otimes \ldots \otimes c(x_n)$ as follows:
\begin{equation}
\begin{split}
& \langle \, \Phi^{\otimes n} \, \rangle
= \langle \, \underbrace{\, \Phi \otimes \Phi \otimes \ldots \otimes \Phi \,}_n \, \rangle \\
& = \int d^d x_1 d^d x_2 \ldots d^d x_n \,
\langle \, \varphi (x_1) \, \varphi (x_2) \, \ldots \, \varphi (x_n) \, \rangle \,
c (x_1) \otimes c(x_2) \otimes \ldots \otimes c(x_n) \,.
\end{split}
\end{equation}
It was shown in~\cite{Konosu:2024dpo} that ${\bm f}$ in~\eqref{f-definition} satisfies
\begin{equation}
( \, {\bf M} +i \hbar \, {\bf U} \, ) \, {\bm f} = 0 \,,
\label{algebraic-Schwinger-Dyson}
\end{equation}
and it follows from this equation that the correlation functions
extracted from ${\bm f}$ satisfy the Schwinger-Dyson equations.
The equation~\eqref{algebraic-Schwinger-Dyson} was called
the algebraic Schwinger-Dyson equation in~\cite{Konosu:2024dpo}.
The relation of~\eqref{algebraic-Schwinger-Dyson} to the definition
of correlation functions in the approach by Costello and Gwilliam
based on factorization algebras~\cite{Costello:2016vjw, Costello:2021jvx}
is explained in appendix~B of~\cite{Konosu:2023pal}.

This formula for correlation functions required us to divide
${\bf M}$ in the free part ${\bf Q}$ and the interaction part ${\bm m}$.
In~\cite{Konosu:2024zrq}, a new form of the formula which does not involve
this division was proposed.
The new formula required us to choose a solution to the equations of motion,
and it was claimed that the formula gives correlation functions
evaluated on the Lefschetz thimble associated with the solution we chose.
When the theory consists of multiple Lefschetz thimbles,
we need to add contributions from those thimbles,
but in this paper we assume that the theory consists
of a single Lefschetz thimble for simplicity.

\section{The LSZ reduction formula}
\label{section-LSZ}
\setcounter{equation}{0}

In quantum field theory,
on-shell scattering amplitudes are obtained from the correlation functions
using the LSZ reduction formula.
In this section we present a description of the LSZ reduction formula
based on quantum $A_\infty$ algebras.
We consider scalar field theories for illustration,
but the discussion can be extended to more general cases.

When we discuss scattering amplitudes, we require the one-point function to vanish.
When the one-point function
\begin{equation}
\Phi_\ast
= \pi_1 \, \frac{1}{{\bf I} +{\bm h} \, {\bm m} +i \hbar \, {\bm h} \, {\bf U}} \, {\bf 1}
\end{equation}
is nonvanishing, we expand the action around $\Phi_\ast$.
The action after the expansion is described by the coderivation $\widetilde{\bf M}$
which is related to the coderivation ${\bf M}$ for the action before the expansion as follows:
\begin{equation}
\widetilde{\bf M} = e^{-{\bf \Phi}_\ast} \, {\bf M} \, e^{{\bf \Phi}_\ast} \,,
\end{equation}
where ${\bf \Phi}_\ast$ is the coderivation associated with $\Phi_\ast$.
See subsection~3.1 of~\cite{Konosu:2024zrq} for more details of this procedure.
In what follows we consider the theory after this expansion,
although we do not add tildes explicitly.

We define the coderivation $\widehat{\bf Q}_\epsilon$ by
\begin{equation}
\widehat{\bf Q}_\epsilon \, \pi_2 \,
\frac{1}{{\bf I} +{\bm h} \, {\bm m} +i \hbar \, {\bm h} \, {\bf U}} \, {\bf 1}
= {}-i \hbar \, \pi_2 \, {\bf U} \, {\bf 1} \,.
\end{equation}
The operator $\widehat{Q}_\epsilon$ is the inverse of the exact propagator.
When we write the two-point function as
\begin{equation}
\pi_2 \, \frac{1}{{\bf I} +\widetilde{\bm h} \, \widetilde{\bm m} +i \hbar \, \widetilde{\bm h} \, {\bf U}} \, {\bf 1}
= {}-i \hbar \int d^d x_1 \, d^d x_2 \int \frac{d^d k}{(2 \pi)^d} \, \frac{e^{-ik \, (x_1-x_2)}}{K(k^2)} \,
c(x_1) \otimes c(x_2) \,,
\end{equation}
$\widehat{Q}_\epsilon$ is given by
\begin{equation}
\widehat{Q}_\epsilon \, c(x) = \int \frac{d^d p}{(2 \pi)^d} \, K(k^2) \, e^{-ik \, (x-y)} \, d(y) \,.
\end{equation}
In terms of the self-energy $\Pi (k^2)$, $K (k^2)$ is given by
\begin{equation}
K (k^2) = k^2 + m^2 -i \epsilon -\Pi (k^2) \,.
\end{equation}
The LSZ reduction formula for the scattering amplitude $\mathcal{S} \, ( \, k_1, k_2, \ldots , k_n \, )$ is then
\begin{equation}
\mathcal{S} \, ( \, k_1, k_2, \ldots , k_n \, )
= \biggl( \, \frac{i}{\hbar} \, \biggr)^n \,
\lim_{\epsilon \to 0} \, \omega_n \, \Bigl( \, \pi_n \,
\frac{1}{{\bf I} +{\bm h} \, {\bm m} +i \hbar \, {\bm h} \, {\bf U}} \, {\bf 1} \,,\,
\widehat{Q}_\epsilon \, \Phi_1 \otimes \widehat{Q}_\epsilon \, \Phi_2 \otimes \ldots
\otimes \widehat{Q}_\epsilon \, \Phi_n \,
\Bigr) \,,
\label{LSZ}
\end{equation}
where $\omega_n$ for the scalar field theory is 
\begin{equation}
\omega_n \, ( \, c (x_1)  \otimes c (x_2) \otimes \ldots \otimes c (x_n) \,,
d (x'_1) \otimes d (x'_2) \otimes \ldots \otimes d (x'_n) \, )
= \prod_{i=1}^n \, \omega \, ( \, c (x_i) \,, d (x'_i) \, ) \,.
\end{equation}
See subsection~4.1 of~\cite{Konosu:2023pal} for the general definition of $\omega_n$
including fermions.
We consider generic cases where all the momenta $k_i$'s are different.
The states $\Phi_i$ are chosen such that
\begin{equation}
\widehat{Q} \, \Phi_i = 0 \,,
\end{equation}
where
\begin{equation}
\widehat{Q} = \lim_{\epsilon \to 0} \, \widehat{Q}_\epsilon \,.
\end{equation}
For the scalar field theory, $\Phi_i$ is given by
\begin{equation}
\Phi_i = \frac{1}{\sqrt{R}} \int d^d x \, e^{ik_i x} \, c(x) \quad
\end{equation}
with
\begin{equation}
k_i^2 = {}-m_{\rm phys}^2 \,,
\end{equation}
where $m_{\rm phys}$ and $R$ are determined by
\begin{equation}
K( {}-m_{\rm phys}^2 \, ) = 0
\end{equation}
and
\begin{equation}
\frac{d K(s)}{ds} \biggr|_{s = -m_{\rm phys}^2} = \frac{1}{R} \,.
\end{equation}

The expression
\begin{equation}
\biggl( \, \frac{i}{\hbar} \, \biggr)^n \,
\lim_{\epsilon \to 0} \, \omega_n \, \Bigl( \, \pi_n \,
\frac{1}{{\bf I} +{\bm h} \, {\bm m} +i \hbar \, {\bm h} \, {\bf U}} \, {\bf 1} \,,\,
\widehat{Q}_\epsilon \, \Phi_1 \otimes \widehat{Q}_\epsilon \, \Phi_2 \otimes \ldots
\otimes \widehat{Q}_\epsilon \, \Phi_n \,
\Bigr)
\label{comparison-LSZ}
\end{equation}
looks rather different from
\begin{equation}
\pi_1 \, {\bf P} \, {\bm m} \,
\frac{1}{{\bf I} +{\bm h} \, {\bm m} +i \hbar \, {\bm h} \, {\bf U}} \, {\bf P} \,.
\label{comparison-minimal-model}
\end{equation}
In~\eqref{comparison-LSZ},
disconnected diagrams are included
and ${\bm h}$ for $P = 0$ is used.
In~\eqref{comparison-minimal-model},
only connected diagrams are included
and ${\bm h}$ for $P \ne 0$ is used.

\section{Connected correlation functions}
\label{section-connected}
\setcounter{equation}{0}

To understand the relation between~\eqref{comparison-LSZ} and ~\eqref{comparison-minimal-model},
we should consider connected correlation functions.
In this section we derive a formula for connected correlation functions
based on quantum $A_\infty$ algebras.

\subsection{The generating functional of connected correlation functions}

It is convenient to consider the generating functional of correlation functions.
For scalar field theories,
the generating functional of correlation functions $Z (J)$ is defined by
\begin{equation}
\begin{split}
\frac{Z (J)}{Z (0)} & = 1 +\frac{i}{\hbar} \int d^d x_1 \, \langle \, \varphi (x_1) \, \rangle \, J (x_1)
+\frac{1}{2!} \, \Bigl( \, \frac{i}{\hbar} \, \Bigr)^2 \int d^d x_1 \, d^d x_2 \,
\langle \, \varphi (x_1) \, \varphi (x_2) \, \rangle \, J (x_1) \, J (x_2) \\
& \quad~
+\frac{1}{3!} \, \Bigl( \, \frac{i}{\hbar} \, \Bigr)^3 \int d^d x_1 \, d^d x_2 \, d^d x_3 \,
\langle \, \varphi (x_1) \, \varphi (x_2) \, \varphi (x_3) \, \rangle \,
J (x_1) \, J (x_2) \, J (x_3) + \, \ldots \,.
\end{split}
\end{equation}
We define
\begin{equation}
\Phi = \int d^d x \, \varphi (x) \, c(x) \,, \qquad
J = \int d^d x \, d(x) \, J (x) \,.
\end{equation}
Since
\begin{equation}
\omega \, ( \, c(x) \,,\, d(x') \, ) = \delta^d (x-x') \,,
\end{equation}
we find
\begin{equation}
\omega \, ( \, \Phi \,,\, J \, ) = \int d^d x \, \varphi (x) \, J (x) \,.
\end{equation}
The generating functional $Z (J)$ can be written as
\begin{equation}
\begin{split}
\frac{Z (J)}{Z (0)} & = 1 +\frac{i}{\hbar} \, \omega \, ( \, \langle \, \Phi \, \rangle \,,\, J \, )
+\frac{1}{2!}  \, \Bigl( \, \frac{i}{\hbar} \, \Bigr)^2
\omega_2 \, ( \, \langle \, \Phi \otimes \Phi \, \rangle \,,\, J \otimes J \, ) \\
& \quad~
+\frac{1}{3!} \, \Bigl( \, \frac{i}{\hbar} \, \Bigr)^3
\omega_3 \, ( \, \langle \, \Phi \otimes \Phi \otimes \Phi \, \rangle \,,\,
J \otimes J \otimes J \, ) + \, \ldots \,.
\end{split}
\end{equation}
Using the formula for correlation functions based on quantum $A_\infty$ algebras,
this is expressed as
\begin{align}
\frac{Z (J)}{Z (0)} & = 1 +\frac{i}{\hbar} \,
\omega \, ( \, \pi_1 \, \frac{1}{{\bf I} +{\bm h} \, {\bm m} +i \hbar \, {\bm h} \, {\bf U}} \, {\bf 1} \,,\, J \, )
+\frac{1}{2!}  \, \Bigl( \, \frac{i}{\hbar} \, \Bigr)^2
\omega_2 \, ( \, \pi_2 \, \frac{1}{{\bf I} +{\bm h} \, {\bm m} +i \hbar \, {\bm h} \, {\bf U}} \, {\bf 1} \,,\,
J \otimes J \, ) \nonumber \\
& \quad~
+\frac{1}{3!} \, \Bigl( \, \frac{i}{\hbar} \, \Bigr)^3
\omega_3 \, ( \, \pi_3 \, \frac{1}{{\bf I} +{\bm h} \, {\bm m} +i \hbar \, {\bm h} \, {\bf U}} \, {\bf 1} \,,\,
J \otimes J \otimes J \, ) + \, \ldots \,.
\end{align}

The generating functional of connected correlation functions $W (J)$ is defined by
\begin{equation}
\begin{split}
\frac{i}{\hbar} \, W (J)
& = \frac{i}{\hbar} \, W (0)
+\frac{i}{\hbar} \int d^d x_1 \, \langle \, \varphi (x_1) \, \rangle_C \, J (x_1) \\
& \quad~
+\frac{1}{2!} \, \Bigl( \, \frac{i}{\hbar} \, \Bigr)^2 \int d^d x_1 \, d^d x_2 \,
\langle \, \varphi (x_1) \, \varphi (x_2) \, \rangle_C \, J (x_1) \, J (x_2) \\
& \quad~
+\frac{1}{3!} \, \Bigl( \, \frac{i}{\hbar} \, \Bigr)^3 \int d^d x_1 \, d^d x_2 \, d^d x_3 \,
\langle \, \varphi (x_1) \, \varphi (x_2) \, \varphi (x_3) \, \rangle_C \,
J (x_1) \, J (x_2) \, J (x_3) + \, \ldots \,.
\end{split}
\end{equation}
To describe connected correlation functions,
let us consider the coproduct
\begin{equation}
\triangle \,
\frac{1}{{\bf I} +{\bm h} \, {\bm m} +i \hbar \, {\bm h} \, {\bf U}} \, {\bf P} \,.
\end{equation}
The calculation of this coproduct is presented in appendix~\ref{appendix-coproduct}.
When $\hbar = 0$, we find
\begin{equation}
\triangle \,
\frac{1}{{\bf I} +{\bm h} \, {\bm m}} \, {\bf P}
= \biggl( \, \frac{1}{{\bf I} +{\bm h} \, {\bm m}} \, {\bf P}
\otimes' \frac{1}{{\bf I} +{\bm h} \, {\bm m} } \, {\bf P} \, \biggr) \, \triangle \,.
\end{equation}
This means that the operator
\begin{equation}
\frac{1}{{\bf I} +{\bm h} \, {\bm m}} \, {\bf P}
\end{equation}
is a cohomomorphism.
When $\hbar \ne 0$, we find
\begin{equation}
\begin{split}
& \triangle \,
\frac{1}{{\bf I} +{\bm h} \, {\bm m} +i \hbar \, {\bm h} \, {\bf U}} \, {\bf P} \\
& = \biggl( \, \frac{1}{{\bf I} +{\bm h} \, {\bm m} +i \hbar \, {\bm h} \, {\bf U}} \, {\bf P}
\otimes' \frac{1}{{\bf I} +{\bm h} \, {\bm m} +i \hbar \, {\bm h} \, {\bf U}} \, {\bf P} \, \biggr) \, \triangle \\
& \quad~ +\sum_{n=1}^\infty \int d^d x_1 \ldots d^d x_n \, \biggl[ \, 
\Bigl( \,
\prod_{i=1}^n {\bm c} (x_i)  \, \Bigr) \,
\frac{1}{{\bf I} +{\bm h} \, {\bm m} +i \hbar \, {\bm h} \, {\bf U}} \, {\bf P} \\
& \quad ~
\otimes'
\prod_{i=1}^n \biggl( \, \frac{1}{{\bf I} +{\bm h} \, {\bm m} +i \hbar \, {\bm h} \, {\bf U}} \, 
( {}-i \hbar \, {\bm h} \, {\bm d} (x_i) \, ) \, \biggr) \,
\frac{1}{{\bf I} +{\bm h} \, {\bm m} +i \hbar \, {\bm h} \, {\bf U}} \, {\bf P} \,
\biggr] \, \triangle \,.
\end{split}
\label{coproduct-formula}
\end{equation}
This means that the operator
\begin{equation}
\frac{1}{{\bf I} +{\bm h} \, {\bm m} +i \hbar \, {\bm h} \, {\bf U}} \, {\bf P}
\end{equation}
is {\it not} a cohomomorphism.
Let us then consider
\begin{equation}
\omega_n \, ( \, \pi_n \, \frac{1}{{\bf I} +{\bm h} \, {\bm m} +i \hbar \, {\bm h} \, {\bf U}} \, {\bf 1} \,,\,
J^{\otimes n} \, )
\end{equation}
in $Z(J)$.
This contains contributions from disconnected diagrams,
and we want to extract contributions from connected diagrams.
We use the identity
\begin{equation}
\pi_n = \bigtriangledown \, ( \, \pi_{n-1} \otimes' \pi_1 \, ) \, \triangle
\end{equation}
to find
\begin{equation}
\begin{split}
& \pi_n \,
\frac{1}{{\bf I} +{\bm h} \, {\bm m} +i \hbar \, {\bm h} \, {\bf U}} \, {\bf 1}
= \bigtriangledown \, ( \, \pi_{n-1} \otimes' \pi_1 \, ) \, \triangle \,
\frac{1}{{\bf I} +{\bm h} \, {\bm m} +i \hbar \, {\bm h} \, {\bf U}} \, {\bf 1} \\
& = \pi_{n-1} \, \frac{1}{{\bf I} +{\bm h} \, {\bm m} +i \hbar \, {\bm h} \, {\bf U}} \, {\bf 1}
\otimes \pi_1 \, \frac{1}{{\bf I} +{\bm h} \, {\bm m} +i \hbar \, {\bm h} \, {\bf U}} \, {\bf 1} \\
& \quad~ +\sum_{k=1}^{n-1} \int d^d x_1 \ldots d^d x_k \, \biggl[ \,
\pi_{n-1} \, \Bigl( \,
\prod_{i=1}^k {\bm c} (x_i)  \, \Bigr)
\frac{1}{{\bf I} +{\bm h} \, {\bm m} +i \hbar \, {\bm h} \, {\bf U}} \, {\bf 1} \\
& \qquad \qquad \quad
\otimes \pi_1 \, \prod_{i=1}^k \biggl( \,
\frac{1}{{\bf I} +{\bm h} \, {\bm m} +i \hbar \, {\bm h} \, {\bf U}} \,
( {}-i \hbar \, {\bm h} \, {\bm d} (x_i) \, ) \, \biggr) \,
\frac{1}{{\bf I} +{\bm h} \, {\bm m} +i \hbar \, {\bm h} \, {\bf U}} \, {\bf 1} \,
\biggr] \,.
\end{split}
\end{equation}
The term
\begin{equation}
\pi_{n-1} \, \frac{1}{{\bf I} +{\bm h} \, {\bm m} +i \hbar \, {\bm h} \, {\bf U}} \, {\bf 1}
\otimes \pi_1 \, \frac{1}{{\bf I} +{\bm h} \, {\bm m} +i \hbar \, {\bm h} \, {\bf U}} \, {\bf 1}
\end{equation}
on the right-hand side clearly corresponds to contributions from disconnected diagrams. 
Let us next consider the term
\begin{equation}
\begin{split}
& \int d^d x_1 \ldots d^d x_k \, \biggl[ \,
\pi_{n-1} \, \Bigl( \,
\prod_{i=1}^k {\bm c} (x_i)  \, \Bigr)
\frac{1}{{\bf I} +{\bm h} \, {\bm m} +i \hbar \, {\bm h} \, {\bf U}} \, {\bf 1} \\
& \qquad \qquad \qquad \quad~~
\otimes \pi_1 \, \prod_{i=1}^k \biggl( \,
\frac{1}{{\bf I} +{\bm h} \, {\bm m} +i \hbar \, {\bm h} \, {\bf U}} \,
( {}-i \hbar \, {\bm h} \, {\bm d} (x_i) \, ) \, \biggr) \,
\frac{1}{{\bf I} +{\bm h} \, {\bm m} +i \hbar \, {\bm h} \, {\bf U}} \, {\bf 1} \,
\biggr] \,.
\end{split}
\end{equation}
The part to the left of the symbol $\otimes$
and the part to the right of the symbol $\otimes$
can be connected through $c (x_i)$ and $d (x_i)$
by the integration over $x_i$.
However, the product of the coderivations
${\bm c} (x_1) \, {\bm c} (x_2) \, \ldots \, {\bm c} (x_k)$
act in the last step so that all of $c (x_1)$, $c (x_2)$, \ldots , and $c (x_k)$
are directly contracted with the sources.
This means that the remaining $n-1-k$ sources to be contracted
with the part to the left of the symbol $\otimes$ cannot be connected
to the part to the right of the symbol $\otimes$.
Therefore, the contributions from connected diagrams
correspond to the term with $k = n-1$:
\begin{equation}
\begin{split}
& \int d^d x_1 \ldots d^d x_{n-1} \, \biggl[ \, \pi_{n-1} \, \Bigl( \,
\prod_{i=1}^{n-1} {\bm c} (x_i) \, \Bigr) \, {\bf 1} \\
& \otimes \pi_1 \, \prod_{i=1}^{n-1} \biggl( \, 
\frac{1}{{\bf I} +{\bm h} \, {\bm m} +i \hbar \, {\bm h} \, {\bf U}} \,
( {}-i \hbar \, {\bm h} \, {\bm d} (x_i) \, )  \, \biggr) \,
\frac{1}{{\bf I} +{\bm h} \, {\bm m} +i \hbar \, {\bm h} \, {\bf U}} \, {\bf 1} \, \biggr] \,.
\end{split}
\end{equation}
Since
\begin{equation}
\omega \, ( \, c(x) \,,\, J \, ) = J (x) \,,
\end{equation}
the contraction of this term with $J^{\otimes n}$ gives
\begin{equation}
\begin{split}
& \int d^d x_1 \ldots d^d x_{n-1} \, \omega_n \, \biggl( \, \pi_{n-1} \, \Bigl( \,
\prod_{i=1}^{n-1} {\bm c} (x_i) \, \Bigr) \, {\bf 1} \\
& \otimes \pi_1 \, \prod_{i=1}^{n-1} \biggl( \,
\frac{1}{{\bf I} +{\bm h} \, {\bm m} +i \hbar \, {\bm h} \, {\bf U}} \,
( {}-i \hbar \, {\bm h} \, {\bm d} (x_i) \, ) \, \biggr) \,
\frac{1}{{\bf I} +{\bm h} \, {\bm m} +i \hbar \, {\bm h} \, {\bf U}} \, {\bf 1} \,,\,
J^{\otimes n} \, \biggr) \\
& = (n-1)! \,\, \omega \, \biggl( \, 
\pi_1 \, \biggl( \, \frac{1}{{\bf I} +{\bm h} \, {\bm m} +i \hbar \, {\bm h} \, {\bf U}} \,
( {}-i \hbar \, {\bm h} \, {\bm J} \, ) \, \biggr)^{n-1}
\frac{1}{{\bf I} +{\bm h} \, {\bm m} +i \hbar \, {\bm h} \, {\bf U}} \, {\bf 1} \,,\,
J \, \biggr) \,.
\end{split}
\end{equation}
The generating functional of connected correlation functions $W (J)$ is thus given by
\begin{equation}
\begin{split}
& \frac{i}{\hbar} \, W (J)
-\frac{i}{\hbar} \, W (0) \\
& = \sum_{n=1}^\infty \frac{1}{n} \, \Bigl( \, \frac{i}{\hbar} \, \Bigr)^n
\omega \, \biggl( \, 
\pi_1 \, \biggl( \, \frac{1}{{\bf I} +{\bm h} \, {\bm m} +i \hbar \, {\bm h} \, {\bf U}} \,
( {}-i \hbar \, {\bm h} \, {\bm J} \, ) \, \biggr)^{n-1}
\frac{1}{{\bf I} +{\bm h} \, {\bm m} +i \hbar \, {\bm h} \, {\bf U}} \, {\bf 1} \,,\,
J \, \biggr) \\
& = \frac{i}{\hbar} \, \sum_{n=1}^\infty \frac{1}{n} \,
\omega \, \biggl( \, 
\pi_1 \, \biggl( \, \frac{1}{{\bf I} +{\bm h} \, {\bm m} +i \hbar \, {\bm h} \, {\bf U}} \,
{\bm h} \, {\bm J} \, \biggr)^{n-1}
\frac{1}{{\bf I} +{\bm h} \, {\bm m} +i \hbar \, {\bm h} \, {\bf U}} \, {\bf 1} \,,\,
J \, \biggr) \,.
\end{split}
\end{equation}
We conclude that
\begin{equation}
\begin{split}
W (J) = W (0)
+\sum_{n=1}^\infty \frac{1}{n} \,
\omega \, \biggl( \, 
\pi_1 \, \biggl( \, \frac{1}{{\bf I} +{\bm h} \, {\bm m} +i \hbar \, {\bm h} \, {\bf U}} \,
{\bm h} \, {\bm J} \, \biggr)^{n-1}
\frac{1}{{\bf I} +{\bm h} \, {\bm m} +i \hbar \, {\bm h} \, {\bf U}} \, {\bf 1} \,,\,
J \, \biggr) \,.
\end{split}
\label{W(J)}
\end{equation}
This can also be expressed as\footnote{
There is a different derivation of $W(J)$ which directly gives
this expression written in terms of an integral over the parameter $t$~\cite{Konosu:2025}.
}
\begin{equation}
W (J) -W (0)
= \int_0^1 dt \, \omega \, \biggl( \, \pi_1 \,
\frac{1}{{\bf I} +{\bm h} \, {\bm m} -t \, {\bm h} \, {\bm J} +i \hbar \, {\bm h} \, {\bf U}} \, {\bf 1} \,,\,
J \, \biggr) \,.
\end{equation}

\subsection{The LSZ reduction formula for connected correlation functions}
\label{subsection-1PI}

In quantum field theory, we know that connected correlation functions
can be expressed in terms of the 1PI effective action.
When we consider perturbation theory,
the description of the theory based on quantum $A_\infty$ algebras
generates the same set of Feynman diagrams
so that connected diagrams can be organized in terms of 1PI diagrams.
We denote the coderivation which describes the 1PI effective action by ${\bf \Gamma}$:
\begin{equation}
{\bf \Gamma} = \sum_{n=0}^\infty {\bf \Gamma}_n \,.
\end{equation}
In a specific theory, we can calculate $\Gamma_n$ from
$m_n$, $h$, and the coderivations in ${\bf U}$
at each order in perturbation theory.
For example, the expressions of $\Gamma_0$ and $\Gamma_1$ for $\varphi^3$ theory at one loop
were calculated in~\cite{Okawa:2022sjf}.
Actually, the expressions of $\Gamma_n$ must be universal
in the description of quantum $A_\infty$ algebras
and do not depend on specific theories,
but we have not yet found the universal expressions.
In this subsection we assume the existence of the universal coderivation ${\bf \Gamma}$
for the 1PI effective action.

Under this assumption,
the generating functional $W(J)$ is expressed in terms of ${\bf \Gamma}$ as
\begin{equation}
W (J) -W (0)
= \sum_{n=1}^\infty \frac{1}{n} \,
\omega \, \biggl( \, 
\pi_1 \, \biggl( \, \frac{1}{{\bf I} +{\bm h} \, {\bf \Gamma}} \,
{\bm h} \, {\bm J} \, \biggr)^{n-1}
\frac{1}{{\bf I} +{\bm h} \, {\bf \Gamma}} \, {\bf 1} \,,\,
J \, \biggr) \,.
\label{W(J)-Gamma}
\end{equation}
When we consider scattering amplitudes, we require that $\Gamma_0 = 0 \,$.
In this case, we have
\begin{equation}
W (J) -W (0)
= \sum_{n=1}^\infty \frac{1}{n} \,
\omega \, \biggl( \, 
\pi_1 \, \biggl( \, \frac{1}{{\bf I} +{\bm h} \, {\bf \Gamma}} \,
{\bm h} \, {\bm J} \, \biggr)^{n-1} {\bf 1} \,,\,
J \, \biggr) \,.
\end{equation}
We then decompose ${\bf \Gamma}$ as
\begin{equation}
{\bf \Gamma} = {\bf \Gamma}_1 + {\bf \Gamma}_{int} \,,
\end{equation}
where ${\bf \Gamma}_1$ describes the self energy.
Since
\begin{equation}
{\bf I} +{\bm h} \, {\bf \Gamma}_1 +{\bm h} \, {\bf \Gamma}_{int}
= ( \, {\bf I} +{\bm h} \, {\bf \Gamma}_1 \, ) \,
\Bigl( \, {\bf I}
+\frac{1}{{\bf I} +{\bm h} \, {\bf \Gamma}_1} \, {\bm h} \, {\bf \Gamma}_{int} \, \Bigr) \,,
\end{equation}
we find
\begin{equation}
\frac{1}{{\bf I} +{\bm h} \, {\bf \Gamma}} \,
{\bm h} \, {\bm J}
= \Bigl( \, {\bf I}
+\frac{1}{{\bf I} +{\bm h} \, {\bf \Gamma}_1} \, {\bm h} \, {\bf \Gamma}_{int} \, \Bigr)^{-1}
\frac{1}{{\bf I} +{\bm h} \, {\bf \Gamma}_1} \, {\bm h} \, {\bm J} \,.
\end{equation}
We define the coderivation ${\bf H}$ for the exact propagator by
\begin{equation}
{\bf H} = \frac{1}{{\bf I} +{\bm h} \, {\bf \Gamma}_1} \, {\bm h} \,,
\end{equation}
and we find
\begin{equation}
\frac{1}{{\bf I} +{\bm h} \, {\bf \Gamma}} \,
{\bm h} \, {\bm J}
= \frac{1}{{\bf I} +{\bf H} \, {\bf \Gamma}_{int}} \,
{\bf H} \, {\bm J} \,.
\end{equation}
The generating functional $W(J)$ is now written as
\begin{equation}
W (J) -W (0)
= \sum_{n=1}^\infty \frac{1}{n} \,
\omega \, \biggl( \, 
\pi_1 \, \biggl( \, \frac{1}{{\bf I} +{\bf H} \, {\bf \Gamma}_{int}} \,
{\bf H} \, {\bm J} \, \biggr)^{n-1} {\bf 1} \,,\,
J \, \biggr) \,.
\end{equation}
Note that
\begin{equation}
\widehat{Q}_\epsilon \, H +H \, \widehat{Q}_\epsilon = \mathbb{I} \,.
\end{equation}
When we apply the LSZ reduction formula
to connected correlation functions expressed in terms of ${\bf H}$ and ${\bf \Gamma}_{int}$,
each $\widehat{Q}_\epsilon$ amputates an external exact propagator.
Internal exact propagators are generically off shell
so that we can safely take the external states to be on shell. 
We can thus calculate scattering amplitudes via the LSZ reduction formula
after taking into account mass renormalization.

\subsection{The operators $W_n$}

Let us come back to the expression
\begin{equation}
\begin{split}
& W (J)-W (0) \\
& = \sum_{n=1}^\infty \frac{1}{n} \,
\omega \, \biggl( \, 
\pi_1 \, \biggl( \, \frac{1}{{\bf I} +{\bm h} \, {\bm m} +i \hbar \, {\bm h} \, {\bf U}} \,
{\bm h} \, {\bm J} \, \biggr)^{n-1}
\frac{1}{{\bf I} +{\bm h} \, {\bm m} +i \hbar \, {\bm h} \, {\bf U}} \, {\bf 1} \,,\,
J \, \biggr) \,.
\end{split}
\end{equation}
Using the notation
\begin{equation}
{\bm f} = \frac{1}{{\bf I} +{\bm h} \, {\bm m} +i \hbar \, {\bm h} \, {\bf U}} \,,
\end{equation}
we write this as
\begin{equation}
\begin{split}
W (J)-W (0)
& = \sum_{n=1}^\infty \frac{1}{n} \,
\omega \, ( \, \pi_1 \, ( \, {\bm f} \, {\bm h} \, {\bm J} \, )^{n-1} \,
{\bm f} \, {\bf 1} \,,\, J \, ) \\
& = \sum_{n=0}^\infty \frac{1}{n+1} \,
\omega \, ( \, \pi_1 \, ( \, {\bm f} \, {\bm h} \, {\bm J} \, )^n \,
{\bm f} \, {\bf 1} \,,\, J \, ) \,.
\end{split}
\end{equation}
This can be further transformed as follows:
\begin{equation}
\begin{split}
W (J)-W (0)
& = \sum_{n=0}^\infty \frac{1}{n+1} \,
\omega \, ( \, \pi_1 \, ( \, {\bm f} \, {\bm h} \, {\bm J} \, )^n \,
{\bm f} \, {\bf 1} \,,\, J \, ) \\
& = \sum_{n=0}^\infty \frac{1}{n+1} \,
\omega \, ( \, \pi_1 \, ( \, {\bm f} \, {\bm h} \, {\bm J} \, )^n \,
{\bm f} \, {\bf 1} \,,\, Q_\epsilon \, hJ \, ) \\
& = \sum_{n=0}^\infty \frac{1}{n+1} \,
\omega \, ( \, hJ \,,\, \pi_1 \, {\bf Q}_\epsilon \, ( \, {\bm f} \, {\bm h} \, {\bm J} \, )^n \,
{\bm f} \, {\bf 1} \, ) \,.
\end{split}
\end{equation}
We write this as
\begin{equation}
W (J) -W (0)
= \sum_{n=0}^\infty \frac{1}{n+1} \,
\omega \, ( \, hJ \,,\,
W_n \, ( hJ )^{\otimes n} \, ) \,.
\end{equation}
Then off-shell amplitudes are defined in terms of $W_n$ by
\begin{equation}
\omega \, ( \, A_1 \,,\,
W_n \, ( A_2 \otimes A_3 \otimes \ldots \otimes A_{n+1} \, ) \, ) \,.
\end{equation}
These correspond to connected diagrams for correlation functions
with the external tree-level propagators amputated.\footnote{
The products $W_n$ at the tree level can be written using a cohomomorphism,
and this description is used to construct
a boundary-string-field-theory-like action in~\cite{Totsuka-Yoshinaka:2025}.
}
In string theory, this is the quantity we obtain 
when we integrate correlation functions of off-shell vertex operators
over the moduli space of Riemann surfaces.
The construction of an action of string field theory
is an inverse problem of constructing string products $m_n$'s
when $W_n$'s are given.

Let us calculate $W_1$, $W_2$, and $W_3$ explicitly.
Since
\begin{equation}
{\bf Q}_\epsilon \, {\bm f} \, {\bm h} \, {\bm J} \, {\bm f} \, {\bf 1}
= {\bf Q}_\epsilon \, {\bm f} \, ( \, {\bm f} \, {\bf 1} \otimes hJ  \, ) \,,
\end{equation}
we find
\begin{equation}
W_1 \, (A_1) = \pi_1 \, {\bf Q}_\epsilon \, {\bm f} \, ( \, {\bm f} \, {\bf 1} \otimes A_1  \, ) \,.
\end{equation}
For $W_2$, it follows from
\begin{equation}
{\bf Q}_\epsilon \, {\bm f} \, {\bm h} \, {\bm J} \, {\bm f} \, {\bm h} \, {\bm J} \, {\bm f} \, {\bf 1}
= {\bf Q}_\epsilon \, {\bm f} \, ( \, {\bm f} \, ( \, {\bm f} \, {\bf 1} \otimes hJ  \, ) \otimes hJ \, ) \,,
\end{equation}
that
\begin{equation}
W_2 \, (A_1 \otimes A_2)
= \pi_1 \, {\bf Q}_\epsilon \, {\bm f} \, ( \, {\bm f} \, ( \, {\bm f} \, {\bf 1} \otimes A_1 \, ) \otimes A_2 \, ) \,.
\end{equation}
For $W_3$, we similarly have
\begin{equation}
{\bf Q}_\epsilon \, {\bm f} \, {\bm h} \, {\bm J} \, {\bm f} \, {\bm h} \, {\bm J} \, 
{\bm f} \, {\bm h} \, {\bm J} \, {\bm f} \, {\bf 1}
= {\bf Q}_\epsilon \, {\bm f} \, ( \, {\bm f} \, ( \, {\bm f} \, ( \, {\bm f} \, {\bf 1}
\otimes hJ  \, ) \otimes hJ \, ) \otimes hJ \, ) \,,
\end{equation}
and we find
\begin{equation}
W_3 \, (A_1 \otimes A_2 \otimes A_3)
= \pi_1 \, {\bf Q}_\epsilon \, {\bm f} \, ( \, {\bm f} \, ( \, {\bm f} \, ( \, {\bm f} \, {\bf 1}
\otimes A_1 \, ) \otimes A_2 \, ) \otimes A_3 \, ) \,.
\end{equation}
It is easy to generalize the calculation to $W_n$,
and we conclude that
\begin{equation}
W_n \, (A_1 \otimes A_2 \otimes \ldots \otimes A_n)
= \pi_1 \, {\bf Q}_\epsilon \,
{\bm f} \, ( \, {\bm f} \, ( \, {\bm f} \, ( \, \ldots
{\bm f} \, ( \, {\bm f} \, {\bf 1} \, \otimes
A_1  \, ) \otimes A_2 \, ) \otimes A_3 \, ) \ldots \otimes A_n \, ) \, ) \,.
\label{W_n}
\end{equation}

\section{The relation to the projection onto on-shell states}
\label{section-relation}
\setcounter{equation}{0}

When we calculate correlation functions,
we use the projection with $P=0$.
The associated contracting homotopy $h$ satisfies
\begin{equation}
Q_\epsilon \, h +h \, Q_\epsilon = \mathbb{I} \,, \quad h^2 = 0 \,,
\end{equation}
and ${\bm h}$ is characterized by
\begin{equation}
\pi_1 \, {\bm h} = h \, \pi_1 \,, \qquad
\triangle \, {\bm h}
= ( \, {\bm h} \otimes' \pi_0 +{\bf I} \otimes' {\bm h} \, ) \, \triangle \,.
\end{equation}
When we calculate on-shell scattering amplitudes at the tree level,
we use the projection onto on-shell states,
and in this case the operator $P$ is nonvanishing.
To distinguish the associated contracting homotopy from $h$,
we denote it by $h_P$ in this section.
The operator $h_P$ satisfies
\begin{equation}
Q \, h_P +h_P \, Q = \mathbb{I} -P \,, \qquad
h_P \, P = 0 \,, \qquad
P \, h_P = 0 \,, \qquad
h_P^2 = 0 \,.
\end{equation}
Corresponding to this, we define ${\bm h}_P$ by
\begin{equation}
\pi_1 \, {\bm h}_P = h_P \, \pi_1 \,, \qquad
\triangle \, {\bm h}_P
= ( \, {\bm h}_P \otimes' {\bf P} +{\bf I} \otimes' {\bm h}_P \, ) \, \triangle \,.
\end{equation}

Let us discuss the relation between
\begin{equation}
\begin{split}
W (J)-W (0)
& = \sum_{n=0}^\infty \frac{1}{n+1} \,
\omega \, ( \, hJ \,,\, \pi_1 \, {\bf Q}_\epsilon \, ( \, {\bm f} \, {\bm h} \, {\bm J} \, )^n \,
{\bm f} \, {\bf 1} \, ) \\
& = \sum_{n=0}^\infty \frac{1}{n+1} \,
\omega \, ( \, hJ \,,\,
W_n \, ( hJ )^{\otimes n} \, )
\end{split}
\end{equation}
and
\begin{equation}
\pi_1 \, {\bf P} \, {\bm m} \,
\frac{1}{{\bf I} +{\bm h}_P \, {\bm m} +i \hbar \, {\bm h}_P \, {\bf U}} \, {\bf P} \,.
\end{equation}
We define $\bar{\bm h}$ by
\begin{equation}
\pi_1 \, \bar{\bm h} = h_P \, \pi_1 \,, \qquad
\triangle \, \bar{\bm h}
= ( \, \bar{\bm h} \otimes' \pi_0 +\mathbb{I} \otimes' \bar{\bm h} \, ) \, \triangle \,.
\end{equation}
For example,
\begin{align}
{\bm h}_P \, \pi_3
& = ( \, h_P \otimes P \otimes P +\mathbb{I} \otimes h_P \otimes P
+\mathbb{I} \otimes \mathbb{I} \otimes h_P \, ) \,
\pi_3 \,, \\
\bar{\bm h} \, \pi_3
& = ( \, \mathbb{I} \otimes \mathbb{I} \otimes h_P \, ) \,
\pi_3 \,, \\
( \, {\bm h}_P -\bar{\bm h} \, ) \, \pi_3
& = ( \, h_P \otimes P \otimes P
+\mathbb{I} \otimes h_P \otimes P \, ) \,
\pi_3 \,.
\end{align}
Note that $h_P$ appears to the right in $\bar{\bm h}$,
and $P$ always appears to the right in ${\bm h}_P -\bar{\bm h}$.
Since $P \, h_P = 0$, this implies that
\begin{equation}
( \, {\bm h}_P -\bar{\bm h} \, ) \, ( \, {\bm m} +i \hbar \, {\bf U} \, ) \,
\bar{\bm h} \, ( \, {\bm m} +i \hbar \, {\bf U} \, ) = 0 \,.
\end{equation}
We thus find that
\begin{equation}
\begin{split}
& \frac{1}{{\bf I} +{\bm h}_P \, ( \, {\bm m} +i \hbar \, {\bf U} \, )}
= \frac{1}{{\bf I} +\bar{\bm h} \, ( \, {\bm m} +i \hbar \, {\bf U} \, )
+( \, {\bm h}_P -\bar{\bm h} \, ) \, ( \, {\bm m} +i \hbar \, {\bf U} \, )} \\
& = \frac{1}{{\bf I} +\bar{\bm h} \, ( \, {\bm m} +i \hbar \, {\bf U} \, )} \,
\frac{1}{{\bf I} +( \, {\bm h}_P -\bar{\bm h} \, ) \, ( \, {\bm m} +i \hbar \, {\bf U} \, )} \,.
\end{split}
\end{equation}
We define
\begin{equation}
{\bm f}_P = \frac{1}{{\bf I} +{\bm h}_P \, {\bm m} +i \hbar \, {\bm h}_P \, {\bf U}} \,, \quad
\bar{\bm f} = \frac{1}{{\bf I} +\bar{\bm h} \, ( \, {\bm m} +i \hbar \, {\bf U} \, )} \,, \quad
{\bm f}'
= \frac{1}{{\bf I} +( \, {\bm h}_P -\bar{\bm h} \, ) \, ( \, {\bm m} +i \hbar \, {\bf U} \, )} \,,
\end{equation}
and then we have
\begin{equation}
{\bm f}_P = \bar{\bm f} \, {\bm f}' \,.
\end{equation}
Since
\begin{equation}
{\bm f}' = \bigtriangledown \, ( \, {\bm f}_P \otimes' P \, \pi_1 \, ) \, \triangle \,,
\end{equation}
we obtain the following recursion relation:
\begin{equation}
{\bm f}_P \, \pi_n = \bar{\bm f} \, {\bm f}' \, \pi_n
= \bar{\bm f} \, \bigtriangledown \, ( \, {\bm f}_P \, \pi_{n-1} \otimes' P \, \pi_1 \, ) \, \triangle \, \pi_n \,.
\end{equation}

Let us consider ${\bm f}_P \, \pi_0$. Since
\begin{equation}
{\bm f}_P \, \pi_0 = \bar{\bm f} \, {\bm f}' \, \pi_0 = \bar{\bm f} \, \pi_0 \,,
\end{equation}
we find
\begin{equation}
{\bm f}_P \, {\bf 1} = \bar{\bm f} \, {\bf 1} \,.
\end{equation}
Let us next consider ${\bm f}_P \, \pi_1$. Since
\begin{equation}
{\bm f} \, \pi_1 = \bar{\bm f}_P \, {\bm f}' \, \pi_1
= \bar{\bm f} \, \bigtriangledown \, ( \, {\bm f}_P \, \pi_0 \otimes' P \, \pi_1 \, ) \, \triangle \, \pi_1
= \bar{\bm f} \, \bigtriangledown \, ( \, \bar{\bm f} \, \pi_0 \otimes' P \, \pi_1 \, ) \, \triangle \, \pi_1 \,,
\end{equation}
we find
\begin{equation}
{\bm f}_P \, A_1
= \bar{\bm f} \, ( \, \bar{\bm f} \, {\bf 1} \otimes P \, A_1 \, ) \,.
\end{equation}
Let us also consider ${\bm f}_P \, \pi_2$. Since
\begin{equation}
\begin{split}
{\bm f}_P \, \pi_2 & = \bar{\bm f} \, {\bm f}' \, \pi_2
= \bar{\bm f} \, \bigtriangledown \, ( \, {\bm f} \, \pi_1 \otimes' P \, \pi_1 \, ) \, \triangle \, \pi_2 \\
& = \bar{\bm f} \, \bigtriangledown \, ( \,
( \, \bar{\bm f} \, \bigtriangledown \, ( \, \bar{\bm f} \, \pi_0 \otimes' P \, \pi_1 \, ) \, \triangle \, \pi_1 \, ) \,
\otimes' P \, \pi_1 \, ) \, \triangle \, \pi_2 \,,
\end{split}
\end{equation}
we find
\begin{equation}
{\bm f}_P \, ( A_1 \otimes A_2 )
= \bar{\bm f} \, ( \, \bar{\bm f} \, ( \, \bar{\bm f} \, {\bf 1} \otimes P \, A_1 \, ) \otimes P \, A_2 \, ) \,.
\end{equation}
It is easy to generalize the calculation to ${\bm f}_P \, \pi_n$,
and we conclude that
\begin{equation}
{\bm f}_P \, (A_1 \otimes A_2 \otimes \ldots \otimes A_n)
= \bar{\bm f} \, ( \, \bar{\bm f} \, ( \, \bar{\bm f} \, ( \, \ldots
\bar{\bm f} \, ( \, \bar{\bm f} \, {\bf 1} \, \otimes
P \, A_1  \, ) \otimes P \, A_2 \, ) \otimes P \, A_3 \, ) \ldots \otimes P \, A_n \, ) \, ) \,.
\label{f_P}
\end{equation}

We note that the structure of $\pi_1 \, {\bm f}_P \, \pi_n$ is similar to that of $W_n$
in~\eqref{W_n}:
\begin{equation}
W_n \, (A_1 \otimes A_2 \otimes \ldots \otimes A_n)
= \pi_1 \, {\bf Q}_\epsilon \,
{\bm f} \, ( \, {\bm f} \, ( \, {\bm f} \, ( \, \ldots
{\bm f} \, ( \, {\bm f} \, {\bf 1} \, \otimes
A_1  \, ) \otimes A_2 \, ) \otimes A_3 \, ) \ldots \otimes A_n \, ) \, ) \,.
\end{equation}
The operator ${\bf Q}_\epsilon$ in $W_n$ simply amputates the final $h$ in ${\bm f}$,
so the operators $\pi_1 \, {\bm f}_P \, \pi_n$ and $W_n$ generate
essentially the same set of Feynman diagrams.
The difference is that
$h_P$ is used in diagrams generated by $\pi_1 \, {\bm f}_P \, \pi_n$
while $h$ is used in diagrams generated by $W_n$.

Let us formally apply the tree-level LSZ reduction formula
to the connected correlation functions.
The relevant quantity is
\begin{equation}
\begin{split}
& \omega \, ( \, h \, Q_\epsilon \, \Phi_1 \,,\,
W_{n-1} \, ( \, h \, Q_\epsilon \, \Phi_2 \otimes \ldots \otimes h \, Q_\epsilon \, \Phi_n \, ) \, ) \\
& = \omega \, ( \, \Phi_1 \,,\,
W_{n-1} \, ( \, \Phi_2 \otimes \ldots \otimes \Phi_n \, ) \, ) \,,
\end{split}
\end{equation}
where we used $Q_\epsilon$ instead of $\widehat{Q}_\epsilon$
because we are considering the tree-level LSZ reduction formula.
This quantity may not be well defined when we take the external states $\Phi_i$
to be on shell at the tree level because of mass renormalization.
We therefore take $\Phi_i$ to be off shell for the moment.
Using the expression of $W_n$ in~\eqref{W_n}, we have
\begin{equation}
\begin{split}
& \omega \, ( \, \Phi_1 \,,\,
W_{n-1} \, ( \, \Phi_2 \otimes \ldots \otimes \Phi_n \, ) \, ) \\
& = \omega \, ( \, \Phi_1 \,,\, \pi_1 \, {\bf Q}_\epsilon \,
{\bm f} \, ( \, {\bm f} \, ( \, \ldots {\bm f} \, ( \, {\bm f} \, {\bf 1} \, \otimes
\Phi_2  \, ) \otimes \Phi_3 \, ) \ldots \otimes \Phi_n \, ) \, ) \,.
\end{split}
\end{equation}
This take the same form as
\begin{equation}
\omega \, ( \, \Phi_1 \,,\, \pi_1 \, {\bf Q}_\epsilon \,
{\bm f}_P \, ( \, \Phi_2  \otimes \Phi_3 \otimes \ldots \otimes \Phi_n \, ) \, )
\end{equation}
with $h_P$ replaced by $h$.
Since
\begin{equation}
( \, {\bf I} +{\bm h} \, {\bm m} +i \hbar \, {\bm h} \, {\bf U} \, ) \,
\frac{1}{{\bf I} +{\bm h} \, {\bm m} +i \hbar \, {\bm h} \, {\bf U}} = {\bf I} \,,
\end{equation}
we have
\begin{equation}
\frac{1}{{\bf I} +{\bm h} \, {\bm m} +i \hbar \, {\bm h} \, {\bf U}}
= {\bf I} -( \, {\bm h} \, {\bm m} +i \hbar \, {\bm h} \, {\bf U} \, ) \,
\frac{1}{{\bf I} +{\bm h} \, {\bm m} +i \hbar \, {\bm h} \, {\bf U}} \,.
\end{equation}
Using this expression, we find
\begin{equation}
\begin{split}
\pi_1 \, {\bf Q}_\epsilon \,
\frac{1}{{\bf I} +{\bm h} \, {\bm m} +i \hbar \, {\bm h} \, {\bf U}}
= Q_\epsilon \, \pi_1
-\pi_1 \, {\bm m} \, \frac{1}{{\bf I} +{\bm h} \, {\bm m} +i \hbar \, {\bm h} \, {\bf U}} \,.
\end{split}
\end{equation}
Therefore,
\begin{equation}
\omega \, ( \, \Phi_1 \,,\, \pi_1 \, {\bf Q}_\epsilon \,
{\bm f}_P \, ( \, \Phi_2  \otimes \Phi_3 \otimes \ldots \otimes \Phi_n \, ) \, )
\end{equation}
with $h_P$ replaced by $h$ is the same as
\begin{equation}
{}-\omega \, ( \, \Phi_1 \,,\, \pi_1 \, {\bm m} \,
{\bm f}_P \, ( \, \Phi_2  \otimes \Phi_3 \otimes \ldots \otimes \Phi_n \, ) \, )
\end{equation}
with $h_P$ replaced by $h$ for $n > 2$.
This proves that the operator $W_n$ with $n > 1$ generates
the same set of Feynman diagrams generated by
\begin{equation}
{}-\pi_1 \, {\bm m} \, \frac{1}{{\bf I} +{\bm h}_P \, {\bm m} +i \hbar \, {\bm h}_P \, {\bf U}} \, \pi_n
\label{quantum-minimal-model}
\end{equation}
with $h_P$ replaced by $h$.
For tree diagrams, internal propagators are generically off shell
when we take the external states to be on shell.
In this case we can safely take the external states to be on shell,
and the operators $h_P$ and $h$ coincide when the momentum is off shell.
Therefore, we conclude that
\begin{equation}
\omega \, ( \, \Phi_1 \,,\, W_{n-1} \, ( \, \Phi_2 \otimes \ldots \otimes \Phi_n \, ) \, ) \,
\biggr|_{\hbar \to 0}
= {}-\omega \, ( \, \Phi_1 \,,\, \pi_1 \, {\bf P} \, {\bm m} \,
\frac{1}{{\bf I} +{\bm h}_P \, {\bm m}} \, {\bf P} \,
( \, \Phi_2 \otimes \ldots \otimes \Phi_n \, ) \, )
\end{equation}
for $n > 2$.

Let us turn to loop diagrams.
The operator $W_n$ with $n > 1$ generates
the same set of Feynman diagrams generated by~\eqref{quantum-minimal-model}
with $h_P$ replaced by $h$,
and this set of diagrams includes loop diagrams.
The evaluation of a loop diagram may not be well defined
when we take the external states to be on shell at the tree level
because of mass renormalization.
For the LSZ reduction formula, we can handle this
by replacing $Q_\epsilon$ with $\widehat{Q}_\epsilon$.
For the operator~\eqref{quantum-minimal-model},
it may seem natural to modify the projection onto the cohomology of $Q$
to the projection onto the cohomology of $\widehat{Q}$.
However, the associated contracting homotopy is $H$ with the projection
onto off-shell momenta, and the operator $h_P$ appearing in~\eqref{quantum-minimal-model}
is not this contracting homotopy.
In quantum field theory, propagators in loops are Feynman propagators,
and this is exactly the case for $W_n$, where $h_P$ is replaced with $h$.
It seems difficult to associate the operator~\eqref{quantum-minimal-model}
with a projection onto a subspace of $\mathcal{H}$ in a simple way.\footnote{
As mentioned in footnote~4 of~\cite{Okawa:2022sjf},
the research on quantum field theory based on quantum $A_\infty$ algebras
initially started with the analysis on loop amplitudes.
However, we encountered this difficulty in making sense of the projection,
which required us a long detour.
}
On the other hand, the LSZ reduction formula provides a well-defined prescription
to loop amplitudes, and we can use this as a basis for further studies. 
It is likely to have an important meaning
that the operator $W_n$ based on the projection with $P=0$
takes formally the same form as~\eqref{quantum-minimal-model}
with $h_P$ replaced by $h$ even when we include loop diagrams,
and we hope to uncover it in future studies.

\section{Conclusions and discussion}
\label{section-conclusions}
\setcounter{equation}{0}

In this paper we presented a well-defined prescription for loop amplitudes
based on quantum $A_\infty$ algebras in~\eqref{LSZ},
which corresponds to the LSZ reduction formula in quantum field theory.
To describe quantum theory in terms of quantum $A_\infty$ algebras,
we use the projection with $P=0$.
For this to be consistent,
the cohomology of the kinetic operator must be trivial,
so we modify $Q$ to $Q_\epsilon$.
The cohomology of $\widehat{Q}_\epsilon$ is also trivial,
and apparently we do not seem to have interesting observables
associated with the cohomology of $\widehat{Q}$.
However, the limit in the LSZ reduction formula~\eqref{LSZ} induces
observables associated with the cohomology of $\widehat{Q}$,
and they physically correspond to on-shell scattering amplitudes.

On the other hand, we know that on-shell scattering amplitudes at the tree level
can be described by the projection onto on-shell states.
To understand the relation between the LSZ reduction formula~\eqref{LSZ}
and this description of on-shell scattering amplitudes,
we derived the expression~\eqref{W(J)} for the generating functional $W(J)$
of connected correlation functions.
This is also one of the main results of this paper.
The generating functional $W(J)$ can also be expressed in terms of the operators $W_n$
given in~\eqref{W_n}.

When $P \ne 0$, the operator ${\bm f}_P$ given by
\begin{equation}
{\bm f}_P = \frac{1}{{\bf I} +{\bm h}_P \, {\bm m} +i \hbar \, {\bm h}_P \, {\bf U}}
\end{equation}
generates a different set of Feynman diagrams
compared to the operator ${\bm f}$ given by
\begin{equation}
{\bm f} = \frac{1}{{\bf I} +{\bm h} \, {\bm m} +i \hbar \, {\bm h} \, {\bf U}}
\end{equation}
associated with the projection with $P=0$.
In section~\ref{section-relation}, we found that the operator $\pi_1 \, {\bm f}_P \, \pi_n$
shares the same structure with $W_n$.
This is highly nontrivial,
and this essentially explains why the LSZ reduction formula and the projection onto on-shell states
give the same scattering amplitudes at the tree level.
For loop amplitudes, we found it difficult to make sense of the projection with $P \ne 0$,
but the LSZ reduction formula~\eqref{LSZ} based on the projection with $P=0$
provides a well-defined prescription
by replacing $Q_\epsilon$ at the tree level with $\widehat{Q}_\epsilon$.

Our claim that the LSZ reduction formula~\eqref{LSZ} gives a well-defined prescription
for loop amplitudes depends on the assumption
that the generating functional $W(J)$~\eqref{W(J)}
can be expressed as~\eqref{W(J)-Gamma} in terms of the coderivation ${\bf \Gamma}$
for the 1PI effective action.
While this is a textbook result in quantum field theory,
this is an intriguing connection between a quantum $A_\infty$ algebra
and a classical $A_\infty$ algebra,
and we want to understand it in a self-contained manner
within the framework of homotopy algebras.
The 1PI effective action also plays a crucial role in renormalization.
It would be therefore important to develop a description of the 1PI effective action
in terms of homotopy algebras.
Furthermore, the description of the 1PI effective action in terms of homotopy algebras
must be universal as we mentioned in subsection~\ref{subsection-1PI}.
In particular, it is universal for both quantum field theory and string field theory.
We want to translate the discussion of mass renormalization
for string theory in~\cite{deLacroix:2017lif}
into the language of homotopy algebras.
For this purpose, $L_\infty$ algebras will be more appropriate.

As its name implies,
the 1PI effective action generates 1PI diagrams in perturbation theory,
and we can in principle calculate the coderivation ${\bf \Gamma}$ at each loop order.
To go beyond perturbation theory, one possibility would be to characterize
the 1PI effective action via the Legendre transformation from $W(J)$
and implement it in the framework of homotopy algebras.
Nonperturbatively, the Legendre transformation can be applied only for convex functions,
and this might be related to the definition of quantum theory
in terms of Lefschetz thimbles.
While the connection of the formula for correlation functions
based on quantum $A_\infty$ algebras to Lefschetz thimbles was uncovered in~\cite{Konosu:2024zrq},
we have not found a principle of how we should combine contributions from multiple
Lefschetz thimbles within the framework of quantum $A_\infty$ algebras.
We hope that the research towards a nonperturbative definition of the 1PI effective action
will give us a clue to this important open problem as well.

\bigskip

\noindent
{\normalfont \bfseries \large Acknowledgments}

\medskip
The results of this paper were presented at the {\it Workshop on String Field Theory and Related Aspects} held in March of 2025.
K.~K, Y.~O. and J.~T.-Y. would like to thank the Institute for Geometry and Physics in Trieste
for hospitality during the workshop
and are grateful to the participants of the workshop for stimulating discussions.
The work of J.~T.-Y. is supported in part by JSPS KAKENHI Grant No.~JP23KJ1311.

\appendix

\section{The calculation of the coproduct~\eqref{coproduct-formula}}
\label{appendix-coproduct}
\setcounter{equation}{0}

We calculate
\begin{equation}
\triangle \,
\frac{1}{{\bf I} +{\bm h} \, {\bm m} +i \hbar \, {\bm h} \, {\bf U}} \, {\bf P}
\end{equation}
in this appendix.
We consider general cases,
and we do not assume that ${\bf P} = \pi_0$.
We denote the basis vector of $\mathcal{H}_1$ by $e^\alpha$,
where $\alpha$ is the label of the basis vectors
which may include both continuous and discrete variables.
The basis vector of $\mathcal{H}_2$ is denoted by $\widetilde{e}_\alpha$,
and repeated indices are implicitly summed over.
The operator ${\bf U}$ is given by
\begin{equation}
{\bf U} = {\bm e}^\alpha \, \widetilde{\bm e}_\alpha \,,
\end{equation}
where ${\bm e}^\alpha$ is the coderivation associated with $e^\alpha$
and $\widetilde{\bm e}_\alpha$ is the coderivation associated with $\widetilde{e}_\alpha$.
The operator ${\bm h}$ satisfies the following relations:
\begin{equation}
{\bm h}^2 = 0 \,, \qquad
{\bf P} \, {\bm h} = 0 \,, \qquad
{\bm h} \, {\bf P} = 0 \,.
\end{equation}
We have
\begin{align}
\triangle \, {\bf P} & = ( \, {\bf P} \otimes' {\bf P} \, ) \, \triangle \,, \\
\triangle \, {\bm h} & = ( \, {\bm h} \otimes' {\bf P} +{\bf I} \otimes' {\bm h} \, ) \, \triangle \,, \\
\triangle \, {\bm m} & = ( \, {\bm m} \otimes' {\bf I} +{\bf I} \otimes' {\bm m} \, ) \, \triangle \,, \\
\triangle \, {\bm e}^\alpha
& = ( \, {\bm e}^\alpha \otimes' {\bf I} +{\bf I} \otimes' {\bm e}^\alpha \, ) \, \triangle \,, \\
\triangle \, \widetilde{\bm e}_\alpha
& = ( \, \widetilde{\bm e}_\alpha \otimes' {\bf I} +{\bf I} \otimes' \widetilde{\bm e}_\alpha \, ) \, \triangle \,.
\end{align}
We then have
\begin{equation}
\triangle \, {\bf U} = ( \, {\bf U} \otimes' {\bf I} +{\bf I} \otimes' {\bf U}
+{\bm e}^\alpha \otimes' \widetilde{\bm e}_\alpha
+\widetilde{\bm e}_\alpha \otimes' {\bm e}^\alpha \, ) \, \triangle \,.
\end{equation}
For ${\bm h} \, {\bm m}$, we have
\begin{equation}
\triangle \, {\bm h} \, {\bm m}
= ( \, {\bm h} \, {\bm m} \otimes' {\bf P}
+{\bm h} \otimes' {\bf P} \, {\bm m}
-{\bm m} \otimes' {\bm h}
+{\bf I} \otimes' {\bm h} \, {\bm m} \, ) \, \triangle \,.
\end{equation}
We further have
\begin{equation}
\triangle \, ( \, {\bm h} \, {\bm m} \, )^n \, {\bf P}
= ( \, {\bm h} \, {\bm m} \otimes' {\bf P}
+{\bf P} \otimes' {\bm h} \, {\bm m} \, )^n \, ( \, {\bf P} \otimes' {\bf P} \, )  \, \triangle \,.
\end{equation}
For ${\bm h} \, {\bf U}$, we have
\begin{equation}
\begin{split}
\triangle \, {\bm h} \, {\bf U}
& = ( \, {\bm h} \, {\bf U} \otimes' {\bf P}
+{\bm h} \otimes' {\bf P} \, {\bf U}
+{\bm h} \, {\bm e}^\alpha \otimes' {\bf P} \, \widetilde{\bm e}_\alpha
+{\bm h} \, \widetilde{\bm e}_\alpha \otimes' {\bf P} \, {\bm e}^\alpha \\
& \quad~
{}-{\bf U} \otimes' {\bm h} +{\bf I} \otimes' {\bm h} \, {\bf U}
+(-1)^{{\bm e}^\alpha} {\bm e}^\alpha \otimes' {\bm h} \, \widetilde{\bm e}_\alpha
+(-1)^{\widetilde{\bm e}_\alpha} \widetilde{\bm e}_\alpha \otimes' {\bm h} \, {\bm e}^\alpha \, ) \, \triangle \,.
\end{split}
\end{equation}
Here and in what follows
$(-1)^{\bm a} = 1$ when ${\bm a}$ is degree even
and $(-1)^{\bm a} = -1$ when ${\bm a}$ is degree odd.
Since $h$ contains $1-P$, we have
\begin{equation}
{\bm h} \, {\bm e}^\alpha \otimes' {\bf P} \, \widetilde{\bm e}_\alpha = 0 \,, \qquad
{\bm h} \, \widetilde{\bm e}_\alpha \otimes' {\bf P} \, {\bm e}^\alpha = 0 \,.
\end{equation}
We thus have
\begin{equation}
\begin{split}
\triangle \, {\bm h} \, {\bf U}
& = ( \, {\bm h} \, {\bf U} \otimes' {\bf P}
+{\bm h} \otimes' {\bf P} \, {\bf U} \\
& \quad~
{}-{\bf U} \otimes' {\bm h} +{\bf I} \otimes' {\bm h} \, {\bf U}
+(-1)^{{\bm e}^\alpha} {\bm e}^\alpha \otimes' {\bm h} \, \widetilde{\bm e}_\alpha
+(-1)^{\widetilde{\bm e}_\alpha} \widetilde{\bm e}_\alpha \otimes' {\bm h} \, {\bm e}^\alpha \, ) \, \triangle \,.
\end{split}
\end{equation}
We further have
\begin{equation}
\triangle \, ( \, {\bm h} \, {\bf U} \, )^n \, {\bf P}
= ( \, {\bm h} \, {\bf U} \otimes' {\bf P}
+{\bf P} \otimes' {\bm h} \, {\bf U}
+(-1)^{{\bm e}^\alpha} {\bm e}^\alpha \otimes' {\bm h} \, \widetilde{\bm e}_\alpha \, )^n \,
( \, {\bf P} \otimes' {\bf P} \, ) \, \triangle \,.
\end{equation}
We combine $\triangle \, {\bm h} \, {\bm m}$ and $\triangle \, {\bm h} \, {\bf U}$ to find
\begin{equation}
\begin{split}
& \triangle \, ( \, {\bm h} \, {\bm m} +i \hbar \, {\bm h} \, {\bf U} \, )^n \, {\bf P} \\
& = ( \, ( \, {\bm h} \, {\bm m} +i \hbar \, {\bm h} \, {\bf U} \, ) \otimes' {\bf P}
+{\bf P} \otimes' ( \, {\bm h} \, {\bm m} +i \hbar \, {\bm h} \, {\bf U} \, )
+(-1)^{{\bm e}^\alpha} i \hbar \, {\bm e}^\alpha \otimes' {\bm h} \, \widetilde{\bm e}_\alpha \, )^n \,
( \, {\bf P} \otimes' {\bf P} \, ) \, \triangle \,.
\end{split}
\end{equation}
Note that
\begin{align}
( \, ( \, {\bm h} \, {\bm m} +i \hbar \, {\bm h} \, {\bf U} \, ) \otimes' {\bf P} \, ) \,
( \, {\bf P} \otimes' ( \, {\bm h} \, {\bm m} +i \hbar \, {\bm h} \, {\bf U} \, ) \, ) & = 0 \,, \\
( \, ( \, {\bm h} \, {\bm m} +i \hbar \, {\bm h} \, {\bf U} \, ) \otimes' {\bf P} \, ) \,
( \, {\bm e}^\alpha \otimes' {\bm h} \, \widetilde{\bm e}_\alpha \, ) & = 0 \,.
\end{align}
Since
\begin{equation}
1+A+B+C = \Bigl( \, 1+A \, \frac{1}{1+B+C} \, \Bigr) \, ( \, 1+B+C \, ) \,,
\end{equation}
we have
\begin{equation}
\frac{1}{1+A+B+C}
= \frac{1}{1+B+C} \, \Bigl( \, 1+A \, \frac{1}{1+B+C} \, \Bigr)^{-1} \,.
\end{equation}
When $A \, B = 0$ and $A \, C = 0 \,$, we find
\begin{equation}
\frac{1}{1+A+B+C}
= \frac{1}{1+B+C} \, \frac{1}{1+A} \,.
\end{equation}
We also use
\begin{equation}
\frac{1}{1+B+C}
= \Bigl( \, 1+\frac{1}{1+B} \, C \, \Bigr)^{-1} \frac{1}{1+B} 
= \sum_{n=0}^\infty  \, \Bigl[ \, \frac{1}{1+B} \, ( {}-C \, ) \, \Bigr]^n \frac{1}{1+B}
\end{equation}
to write
\begin{equation}
\frac{1}{1+A+B+C}
= \sum_{n=0}^\infty \, \Bigl[ \, \frac{1}{1+B} \, ( {}-C \, ) \, \Bigr]^n \, \frac{1}{1+B} \, \frac{1}{1+A} \,.
\end{equation}
Using this, we conclude that
\begin{equation}
\begin{split}
& \triangle \,
\frac{1}{{\bf I} +{\bm h} \, {\bm m} +i \hbar \, {\bm h} \, {\bf U}} \, {\bf P} \\
& = \biggl( \, \frac{1}{{\bf I} +{\bm h} \, {\bm m} +i \hbar \, {\bm h} \, {\bf U}} \, {\bf P}
\otimes' \frac{1}{{\bf I} +{\bm h} \, {\bm m} +i \hbar \, {\bm h} \, {\bf U}} \, {\bf P} \, \biggr) \, \triangle \\
& \quad~ +\sum_{n=1}^\infty \biggl[ \, \Bigl( \,
\prod_{i=1}^n {\bm e}^{\alpha_i} \, \Bigr)
\frac{1}{{\bf I} +{\bm h} \, {\bm m} +i \hbar \, {\bm h} \, {\bf U}} \, {\bf P} \\
& \quad ~
\otimes'
\prod_{i=1}^n \biggl( \, \frac{1}{{\bf I} +{\bm h} \, {\bm m} +i \hbar \, {\bm h} \, {\bf U}} \, 
( {}-(-1)^{{\bm e}^{\alpha_i}} i \hbar \, {\bm h} \, \widetilde{\bm e}_{\alpha_i} \, ) \, \biggr) \,
\frac{1}{{\bf I} +{\bm h} \, {\bm m} +i \hbar \, {\bm h} \, {\bf U}} \, {\bf P} \,
\biggr] \, \triangle \,.
\end{split}
\end{equation}


\small

\begin{thebibliography}{99}

\bibitem{Stasheff:I}
  J.~D.~Stasheff,
  ``Homotopy associativity of $H$-spaces. I,''
  Trans. Amer. Math. Soc. {\bf 108}, 275 (1963).

\bibitem{Stasheff:II}
  J.~D.~Stasheff,   ``Homotopy associativity of $H$-spaces. II,''
  Trans. Amer. Math. Soc. {\bf 108}, 293 (1963).

\bibitem{Getzler-Jones}
  E.~Getzler and J.~D.~S.~Jones,
  ``$A_\infty$-algebras and the cyclic bar complex,''
  Illinois~J.~Math {\bf 34}, 256 (1990).

\bibitem{Markl}
  M.~Markl,
  ``A cohomology theory for $A (m)$-algebras and applications,''
  J. Pure Appl. Algebra {\bf 83}, 141 (1992).

\bibitem{Penkava:1994mu}
  M.~Penkava and A.~S.~Schwarz,
  ``$A_\infty$ algebras and the cohomology of moduli spaces,''
  Trans. Amer. Math. Soc. {\bf 169}, 91 (1995)
  [hep-th/9408064].

\bibitem{Gaberdiel:1997ia}
  M.~R.~Gaberdiel and B.~Zwiebach,
  ``Tensor constructions of open string theories. 1: Foundations,''
  Nucl. Phys. {\bf B505}, 569 (1997)
  [hep-th/9705038].

\bibitem{Zwiebach:1992ie}
B.~Zwiebach,
``Closed string field theory: Quantum action and the Batalin-Vilkovisky master equation,''
Nucl. Phys. B \textbf{390}, 33-152 (1993)
[arXiv:hep-th/9206084 [hep-th]].

\bibitem{Markl:1997bj}
M.~Markl,
``Loop homotopy algebras in closed string field theory,''
Commun. Math. Phys. \textbf{221}, 367-384 (2001)
[arXiv:hep-th/9711045 [hep-th]].

\bibitem{Hohm:2017pnh}
O.~Hohm and B.~Zwiebach,
``$L_{\infty}$ Algebras and Field Theory,''
Fortsch. Phys. \textbf{65}, no.3-4, 1700014 (2017)
[arXiv:1701.08824 [hep-th]].

\bibitem{Jurco:2018sby}
B.~Jur\v{c}o, L.~Raspollini, C.~S\"amann and M.~Wolf,
``$L_\infty$-Algebras of Classical Field Theories and the Batalin-Vilkovisky Formalism,''
Fortsch. Phys. \textbf{67}, no.7, 1900025 (2019)
[arXiv:1809.09899 [hep-th]].

\bibitem{Nutzi:2018vkl}
A.~N\"utzi and M.~Reiterer,
``Amplitudes in YM and GR as a Minimal Model and Recursive Characterization,''
Commun. Math. Phys. \textbf{392}, no.2, 427-482 (2022)
[arXiv:1812.06454 [math-ph]].

\bibitem{Arvanitakis:2019ald}
A.~S.~Arvanitakis,
``The $L_\infty$-algebra of the S-matrix,''
JHEP \textbf{07}, 115 (2019)
[arXiv:1903.05643 [hep-th]].

\bibitem{Macrelli:2019afx}
T.~Macrelli, C.~S\"amann and M.~Wolf,
``Scattering amplitude recursion relations in Batalin-Vilkovisky\textendash{}quantizable theories,''
Phys. Rev. D \textbf{100}, no.4, 045017 (2019)
[arXiv:1903.05713 [hep-th]].

\bibitem{Jurco:2019yfd}
B.~Jur\v{c}o, T.~Macrelli, C.~S\"amann and M.~Wolf,
``Loop Amplitudes and Quantum Homotopy Algebras,''
JHEP \textbf{07}, 003 (2020)
[arXiv:1912.06695 [hep-th]].

\bibitem{Saemann:2020oyz}
C.~Saemann and E.~Sfinarolakis,
``Symmetry Factors of Feynman Diagrams and the Homological Perturbation Lemma,''
JHEP \textbf{12}, 088 (2020)
[arXiv:2009.12616 [hep-th]].

\bibitem{Bonezzi:2023xhn}
R.~Bonezzi, C.~Chiaffrino, F.~Diaz-Jaramillo and O.~Hohm,
``Tree-level Scattering Amplitudes via Homotopy Transfer,''
[arXiv:2312.09306 [hep-th]].

\bibitem{Pius:2013sca}
R.~Pius, A.~Rudra and A.~Sen,
``Mass Renormalization in String Theory: Special States,''
JHEP \textbf{07}, 058 (2014)
[arXiv:1311.1257 [hep-th]].

\bibitem{Pius:2014iaa}
R.~Pius, A.~Rudra and A.~Sen,
``Mass Renormalization in String Theory: General States,''
JHEP \textbf{07}, 062 (2014)
[arXiv:1401.7014 [hep-th]].

\bibitem{Pius:2014gza}
R.~Pius, A.~Rudra and A.~Sen,
``String Perturbation Theory Around Dynamically Shifted Vacuum,''
JHEP \textbf{10}, 070 (2014)
[arXiv:1404.6254 [hep-th]].

\bibitem{Sen:2014dqa}
A.~Sen,
``Gauge Invariant 1PI Effective Action for Superstring Field Theory,''
JHEP \textbf{06}, 022 (2015)
[arXiv:1411.7478 [hep-th]].

\bibitem{Sen:2015hha}
A.~Sen,
``Gauge Invariant 1PI Effective Superstring Field Theory: Inclusion of the Ramond Sector,''
JHEP \textbf{08}, 025 (2015)
[arXiv:1501.00988 [hep-th]].

\bibitem{Sen:2015uoa}
A.~Sen,
``Supersymmetry Restoration in Superstring Perturbation Theory,''
JHEP \textbf{12}, 075 (2015)
[arXiv:1508.02481 [hep-th]].

\bibitem{deLacroix:2017lif}
C.~de Lacroix, H.~Erbin, S.~P.~Kashyap, A.~Sen and M.~Verma,
``Closed Superstring Field Theory and its Applications,''
Int. J. Mod. Phys. A \textbf{32}, no.28n29, 1730021 (2017)
[arXiv:1703.06410 [hep-th]].

\bibitem{Sen:2024nfd}
A.~Sen and B.~Zwiebach,
``String Field Theory: A Review,''
[arXiv:2405.19421 [hep-th]].

\bibitem{Okawa:2022sjf}
Y.~Okawa,
``Correlation functions of scalar field theories from homotopy algebras,''
JHEP \textbf{05}, 040 (2024)
[arXiv:2203.05366 [hep-th]].

\bibitem{Konosu:2023pal}
K.~Konosu and Y.~Okawa,
``Correlation Functions Involving Dirac Fields from Homotopy Algebras I: The Free Theory,''
PTEP \textbf{2025}, no.3, 033B10 (2025)
[arXiv:2305.11634 [hep-th]].

\bibitem{Konosu:2023rkm}
K.~Konosu,
``Correlation Functions Involving Dirac Fields from Homotopy Algebras II: The Interacting Theory,''
PTEP \textbf{2024}, no.9, 093B01 (2024)
[arXiv:2305.13103 [hep-th]].

\bibitem{Konosu:2024dpo}
K.~Konosu and J.~Totsuka-Yoshinaka,
``Noether\textquoteright{}s theorem and Ward-Takahashi identities from homotopy algebras,''
JHEP \textbf{09}, 048 (2024)
[arXiv:2405.09243 [hep-th]].

\bibitem{Konosu:2024zrq}
K.~Konosu and Y.~Okawa,
``Nonperturbative correlation functions from homotopy algebras,''
JHEP \textbf{01}, 152 (2025)
[arXiv:2405.10935 [hep-th]].

\bibitem{Erler:2015uba}
T.~Erler,
``Relating Berkovits and A$_\infty$ superstring field theories; small Hilbert space perspective,''
JHEP \textbf{10}, 157 (2015)
[arXiv:1505.02069 [hep-th]].

\bibitem{Kajiura:2003ax}
H.~Kajiura,
``Noncommutative homotopy algebras associated with open strings,''
Rev. Math. Phys. \textbf{19}, 1-99 (2007)
[arXiv:math/0306332 [math.QA]].

\bibitem{Erler:2020beb}
T.~Erler and H.~Matsunaga,
``Mapping between Witten and lightcone string field theories,''
JHEP \textbf{11}, 208 (2021)
[arXiv:2012.09521 [hep-th]].

\bibitem{Sen:2016qap}
A.~Sen,
``Wilsonian Effective Action of Superstring Theory,''
JHEP \textbf{01}, 108 (2017)
[arXiv:1609.00459 [hep-th]].

\bibitem{Erbin:2020eyc}
H.~Erbin, C.~Maccaferri, M.~Schnabl and J.~Vo\v{s}mera,
``Classical algebraic structures in string theory effective actions,''
JHEP \textbf{11}, 123 (2020)
[arXiv:2006.16270 [hep-th]].

\bibitem{Koyama:2020qfb}
D.~Koyama, Y.~Okawa and N.~Suzuki,
``Gauge-invariant operators of open bosonic string field theory in the low-energy limit,''
[arXiv:2006.16710 [hep-th]].

\bibitem{Arvanitakis:2020rrk}
A.~S.~Arvanitakis, O.~Hohm, C.~Hull and V.~Lekeu,
``Homotopy Transfer and Effective Field Theory I: Tree-level,''
Fortsch. Phys. \textbf{70}, no.2-3, 2200003 (2022)
[arXiv:2007.07942 [hep-th]].

\bibitem{Arvanitakis:2021ecw}
A.~S.~Arvanitakis, O.~Hohm, C.~Hull and V.~Lekeu,
``Homotopy Transfer and Effective Field Theory II: Strings and Double Field Theory,''
Fortsch. Phys. \textbf{70}, no.2-3, 2200004 (2022)
[arXiv:2106.08343 [hep-th]].

\bibitem{Costello:2016vjw}
K.~Costello and O.~Gwilliam,
``Factorization Algebras in Quantum Field Theory: Volume 1,''
Cambridge University Press (2016).

\bibitem{Costello:2021jvx}
K.~Costello and O.~Gwilliam,
``Factorization Algebras in Quantum Field Theory: Volume 2,''
Cambridge University Press (2021).

\bibitem{Konosu:2025}
K.~Konosu and Y.~Okawa, {\it in preparation}.

\bibitem{Totsuka-Yoshinaka:2025}
J.~Totsuka-Yoshinaka,
``BSFT-like action from cohomomorphism,''
{\it to appear}.

\end{thebibliography}
\end{document}